\documentclass[%
 reprint,
 amsmath,amssymb,
 aps,
]{revtex4-2}

\usepackage{graphicx}
\usepackage{dcolumn}
\usepackage{bm}

\begin{document}

\preprint{APS/123-QED}

\title{Self-similarity of temporal interaction networks arises from \\ hyperbolic geometry with time-varying curvature}

\author{Subhabrata Dutta}
\affiliation{Technische Universität Darmstad, Germany}
 
\author{Dipankar Das}%
\affiliation{%
 Jadavpur University, India
}%

\author{Tanmoy Chakraborty}
 \email{Corresponding author. Email: tanchak@iitd.ac.in}
\affiliation{ Indian Institute of Technology Delhi, India
}%

\begin{abstract}
The self-similarity of complex systems has been studied intensely across different domains \cite{general-self-similarity-1, general-self-similarity-2, temporal-scale-2, general-self-similarity-4, general-self-similarity-5, self-similar-2005} due to its potential applications in system modeling, complexity analysis, etc., as well as for deep theoretical interest. Existing studies rely on scale transformations conceptualized over either a definite geometric structure of the system~\cite{self-similar-2005, box-cover-1, general-self-similarity-3,gallos2007review} (very often realized as length-scale transformations) or purely temporal scale transformations~\cite{temporal-scale-1, temporal-scale-2}. However, many physical and social systems are observed as temporal interactions among agents without any definitive geometry. Yet, one can imagine the existence of an underlying notion of distance as the interactions are mostly localized. Analysing only the time-scale transformations over such systems would uncover only a limited aspect of the complexity. In this work, we propose a novel technique of scale transformation that dissects temporal interaction networks under spatio-temporal scales, namely, flow scales. Upon experimenting with multiple social and biological interaction networks, we find that many of them possess a finite fractal dimension under flow-scale transformation. Finally, we relate the emergence of flow-scale self-similarity to the latent geometry of such networks. We observe strong evidence that justifies the assumption of an underlying, variable-curvature hyperbolic geometry that induces self-similarity of temporal interaction networks. Our work bears implications for modeling temporal interaction networks at different scales and uncovering their latent geometric structures.
\end{abstract}

\maketitle

\section{Introduction}
Scale invariance is defined as the characteristic of a dynamical system when its topological/dynamic properties remain the same at different scales~\cite{scale-invariance-seminal-1, scale-invariance-seminal-2,cohen2004fractal, scale-invariance-seminal-3,guimera2003self} such as length, time, and size. Precisely, if one represents such properties of the system as a function $f(s)$ of (spatial/temporal) scale $s$, then for a scale-invariant system, we have $f(\lambda s) = C(\lambda) f(s)$ for some arbitrary constant $\lambda$ and $C(\lambda)$ being scale-independent component~\cite{scale-invariance-definition}. In the case of discrete scale transformations, scale invariance is also identified as self-similarity, i.e., the system under consideration consists of repeating patterns under different length-scales. Certain real-world complex networks have been shown to exhibit self-similarity~\cite{self-similar-2005, self-similar-complex-networks-2, self-similar-complex-networks-3, self-similarity-metric-space-1,kim2007fractality,tsybakov1998self}. A popular approach to testing this property of networks is the {\em box-counting} method~\cite{self-similar-2005, box-cover-1, box-cover-2, box-cover-3}. Conceptually, the size of the box parameterizes the length-scale transformation of the networks; for a self-similar network, the number of boxes needed to cover the network is inversely proportional to a constant power of the size of the box. 

\begin{figure*}[!ht]
    \centering
    \includegraphics[width=\textwidth]{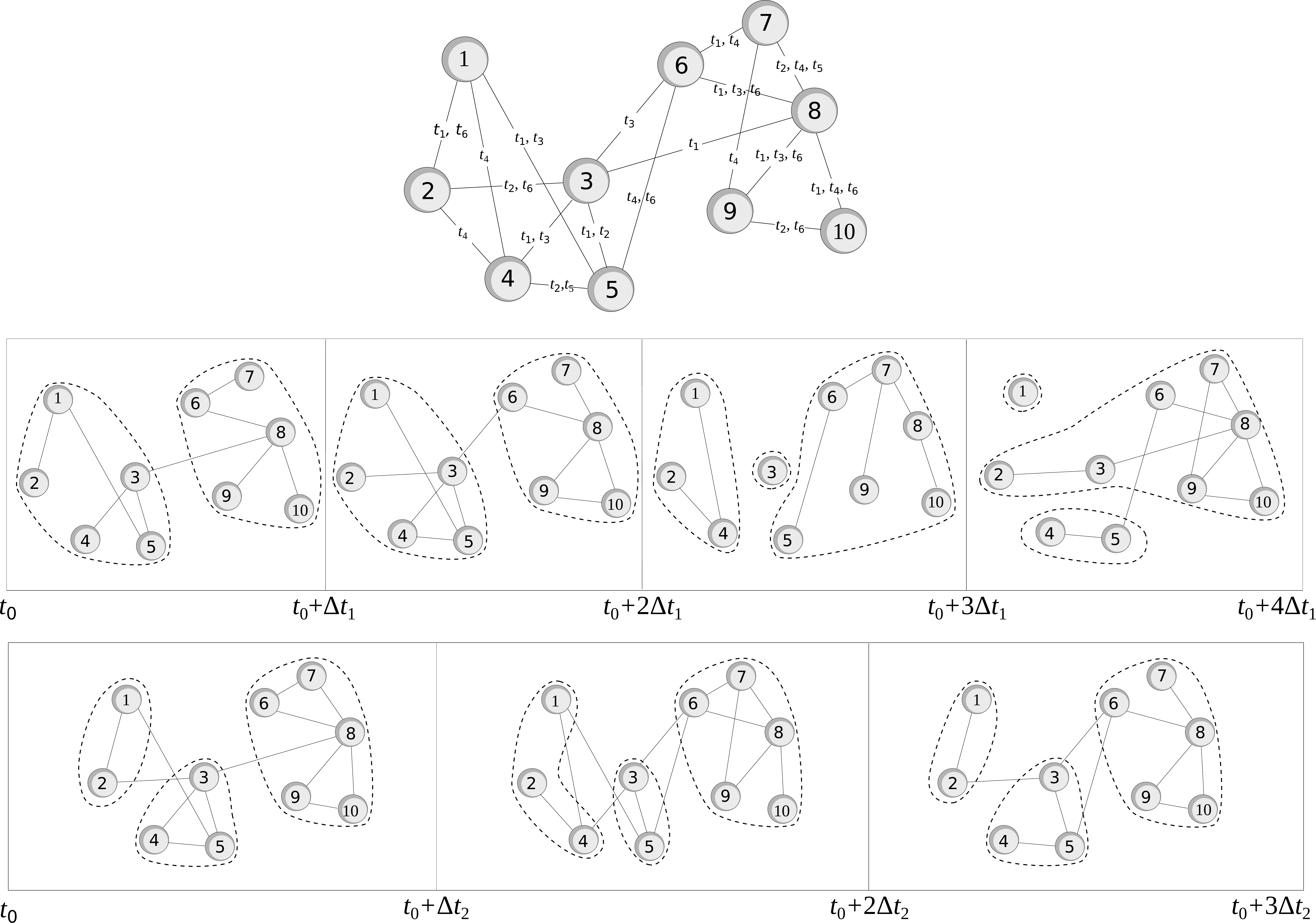}
    \caption{{\bf Schematic description of flow-scale transformation on example temporal interaction network.} {\bf Top:} A temporal interaction network with timestamped interaction edges; here $t_0<t_1<\cdots <t_6$. A snapshot of the network within the time interval $(t_i, t_j)$ would contain edges corresponding to interactions within that interval. {\bf Middle:} Box-covering of the network with box size $=3$ and time size $=\Delta t_1$; here $t_0<t_1\leq t_0+\Delta t_1$, $t_0+\Delta t_1<t_2<t_3\leq t_0+2\Delta t_1$, $t_0+2\Delta t_1<t_4\leq t_0+3\Delta t_1$, and $t_0+3\Delta t_1<t_5<t_6\leq t_0+4\Delta t_1$. In this case, the total number of boxes covering the network is $10$. Similarly in the {\bf bottom}, a box covering with box size $2$ and time size $\Delta t_2$ is shown, where $t_0<t_1<t_2\leq t_0+\Delta t_2$, $t_0+\Delta t_2<t_3<t_4\leq t_0+2\Delta t_2$, and, $t_0+2\Delta t_2<t_5<t_6\leq t_0+3\Delta t_2$. The required number of boxes is $9$.}
    \label{fig:temporal-box-cover}
\end{figure*}

However, these existing methods of scale transformations become limited once we go beyond the static network regime. For a dynamically evolving network, one should consider the time-scale alongside the length-scale (i.e., the temporal evolution of topological properties of the network under different time-scales). The problem is even more challenging in the case of temporal interaction networks (e.g., protein-protein interactions, social interactions over online or physical platforms, etc.). Unlike networks like WWW or routing networks, where there are temporally finite connections between nodes (and, as a result, a tangible geometry over which one can define the idea of length-scale), in the case of interaction networks, the connection between any two nodes is a momentary event. As shown in Figure~\ref{fig:temporal-box-cover}, the geometry of the network quickly changes depending upon the choice of time-scale. In essence, the length- and time-scales become inseparable when discussing the scale transformation and scale invariance of such networks.

Here we propose a novel spatio-temporal box-counting method that analyzes temporal interaction networks under simultaneous length- and time-scale transformations, which we define as {\em flow-scale} transformation. We find that interaction networks can be either scale-invariant, fractal, or non-fractal under such a transformation. We relate this to an underlying hyperbolic geometry; erratically moving random particles over a hyperbolic surface is a fractal object under flow-scale transformation. Furthermore, we empirically observe that a scale-invariant or self-similar point-particle motion can only be observed when the negative curvature of the underlying hyperboloid increases exponentially over time. This expands upon the previous works~\cite{Hyperbolic-graph-embeddings-1, Hyperbolic-graph-embeddings-2, hyperbolic-graph-embeddings-3, hyperbolic-graph-embeddings-4} of embedding static hierarchical network structures into constant curvature hyperbolic geometry, which fails to exhibit self-similar behavior when the temporal regime is taken into account.

\begin{table*}[!ht]
    \centering
    \caption{{\bf Statistics of the real-world interaction networks used in this study.} Except from {\bf DPPIN-Babu}, all other networks have a higher temporal duration. We select a window from them for ease in flow-box computation and visualization. $|\mathcal{V}|$ and $|\mathcal{E}|$ denote the number of nodes and edges, respectively.}
    \begin{tabular}{l|l|r|r|r}
    \hline
        \multicolumn{1}{c|}{Name} & \multicolumn{1}{c|}{Type of interaction} & \multicolumn{1}{c|}{$|\mathcal{V}|$} & \multicolumn{1}{c|}{$|\mathcal{E}|$} & \multicolumn{1}{c}{Duration}\\\hline
        {\tt ia-email} \cite{enron-mail}  & Email interactions & 56931 & 762462 & 720 days\\
        {\tt reddit-hyperlink} 
~\cite{reddit} & Interaction among subreddits via users posting hyperlinks & 33870 & 299016 & 720 days\\
        {\tt DPPIN-Babu}~\cite{DPPIN} & Protein-protein interaction & 4924 & 111466 & 36 units \\
        {\tt ca-cit}~\cite{nr} & Arxiv High Energy Physics paper citation network & 16763 & 2268171 & 2160 days\\
        {\tt superuser}~\cite{superuser} & Comments, questions, and answers on Super User & 188992 & 1309161 & 2160 days \\
        {\tt wiki-talk}~\cite{superuser} & Talkpage interactions among Wikipedia editors & 45443 & 468798 & 990 days\\
        \hline
    \end{tabular}
    \label{tab:real-networks}
\end{table*}

\begin{figure*}[!ht]
    \centering
    \includegraphics[width=\textwidth]{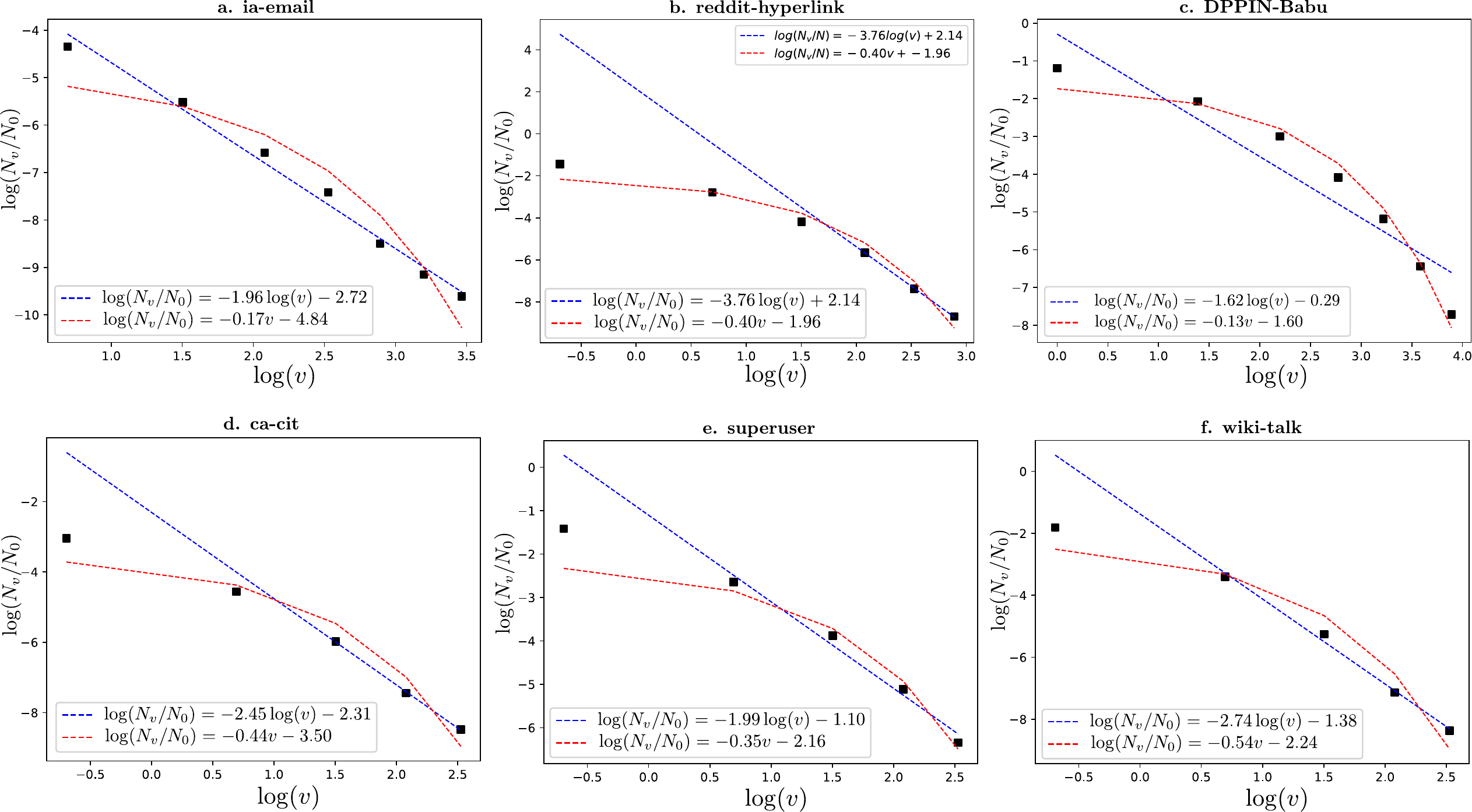}
    \caption{{\bf Scale transformation characteristics of the real temporal interaction networks.} We show the values of $\log(N_v/N_0)$ vs. $\log(v)$ for different real networks. To check for the finiteness of the fractal dimension $d_v$, we fit two regression lines: one assuming a linear relationship between the last four data points of $\log(N_v/N_0)$ vs $\log(v)$ (blue dashed line), another assuming the same between $\log(N_v/N_0)$ vs $v$ (red dashed line). Evidently, the {\tt ia-email} network exhibits scale-invariance regarding Equation~\ref{eq:temporal-fractal-dim}, with a fractal dimension of $1.96$. For {\tt DPPIN-Babu} and {\tt superuser} networks, $d_v$ does not show any bounded variation. The rest of the networks, though have finite fractal dimensions, do not show scale-invariance.}
    \label{fig:real-networks}
\end{figure*}

\section{Results}
\subsection{Discrete scale transformation of temporal networks}
For static networks, box-covering provides the most popular approach to test discrete scale invariance. Given a network $\mathcal{G}=\{\mathcal{V}, \mathcal{E}\}$, a box-cover of size $l_b$ can be defined as $C_{l_b}(\mathcal{G}) = \{c_i|\bigcup_i c_i = V, c_i\cap c_j=\phi\}$, such that, the minimum distance between any two nodes in $c_i$ is at most $l_b$. The fractal dimension of the network is defined as,
\begin{equation}
\label{eq:static-fractal-dim}
    d_b = \lim_{\epsilon\to 0} \frac{\log(N_b/N)}{\log(\epsilon)}
\end{equation}
where $N$ is the number of nodes in the network, $|\mathcal{V}|$; $N_b$ is the number of boxes of size $l_b$ required to cover the network, $|C_{l_b}(\mathcal{G})|$, and $\epsilon=\frac{1}{l_b}$. A network is {\em fractal} if $d$ is finite. Furthermore, if $d_b$ remains constant as $l_b$ varies, then the network is {\em length-scale invariant} or {\em self-similar}. The term $N_b/N$ typically defines the average number of nodes in each box, or its {\em mass}; for a scale-invariant network, the mass of a box follows a power-law variation with its size~\cite{self-similar-2005, box-cover-1}.

Temporal interaction networks naturally extend the spatial scale transformation to the temporal domain, yet it remains unexplored in the literature. A temporal interaction network can be formalized as $\mathcal{G}'=\{\mathcal{V}', \mathcal{E}'\}$, where each edge $e_k\in \mathcal{E}'$ is a triplet $(v_i, v_j, t_k)$ for $v_i,v_j\in \mathcal{V}'$, and $t_k\in [t_\text{start}, t_\text{end}]$ is the timestamp of the corresponding interaction. A popular approach of temporal graph learning is to represent $\mathcal{G}'$ as a sequence of $M$ static snapshots $[G_0, G_1, \cdots, G_M]$ by sampling interactions with a given time window $\Delta t$; each $G_m=\{V_m\subseteq \mathcal{V}', E_m\subseteq \mathcal{E}'\}$ is constructed such that for any $e_k=(v_i, v_j, t_k)\in \mathcal{E}'$, if $t_k \in [t_\text{start} +(m-1)\Delta t, t_\text{start} +m\Delta t]$, $v_i,v_j$ are added to $V_m$ and the edge $(v_i, v_j)$ is added to $E_m$. As shown in Figure~\ref{fig:temporal-box-cover}, the size of the sampling window dictates the granularity of approximation: the smaller the window, the closer the snapshots are to the actual interaction network. Therefore, the sampling time window size $\Delta t$ provides a temporal analogue of the box size $l_b$. Putting them together, we define the modified box counting for a temporal interaction network $\mathcal{G}'$. For a given box size $l_b$ and a sampling time size $\Delta t$, the box-cover is defined as:
\begin{equation}
\label{eq:temporal-scale-transform}
    C_{l_b, \Delta t}(\mathcal{G}') = \bigcup_{m\in [1, M]} C_{l_b}(G_m)
\end{equation}
For brevity, we will denote $|C_{l_b, \Delta t}(\mathcal{G}')|$ as $N_{l_b,\Delta t}$ henceforth. Furthermore, we define the total number of vertices in the span $[t_\text{start}, t_\text{end}]$ for time-window size $\Delta t$ as $N_{0, \Delta t} = \sum_m \lvert V_m \rvert$. In Figure~\ref{fig:temporal-box-cover}, we show a working example. For the given network, $N_{3,\Delta t_1}=10$, whereas $N_{2,\Delta t_2}=9$.

Therefore, if we define the volume of the spatio-temporal boxes as $v=l_b\Delta t$, then one can redefine Equation~\ref{eq:static-fractal-dim} in case of temporal networks as,
\begin{equation}
\label{eq:temporal-fractal-dim}
    d_v = \lim_{v\to \infty} \frac{\log(N_v/N_{0, \Delta t})}{-\log(v)}
\end{equation}

Similar to its static counterpart, a temporal network is scale-invariant if $d_v$ is constant for all scale sizes $v$, and a fractal for a finite limiting value of $d_v$. For a temporal scale size of $\Delta t$, the term $N_v/N_{0, \Delta t}$ again defines the number of nodes within scale volume $v$. A detailed description of flow-scale transformation is presented in {\em Methods}.

\begin{figure*}[!ht]
    \centering
    \includegraphics[width=0.9\textwidth]{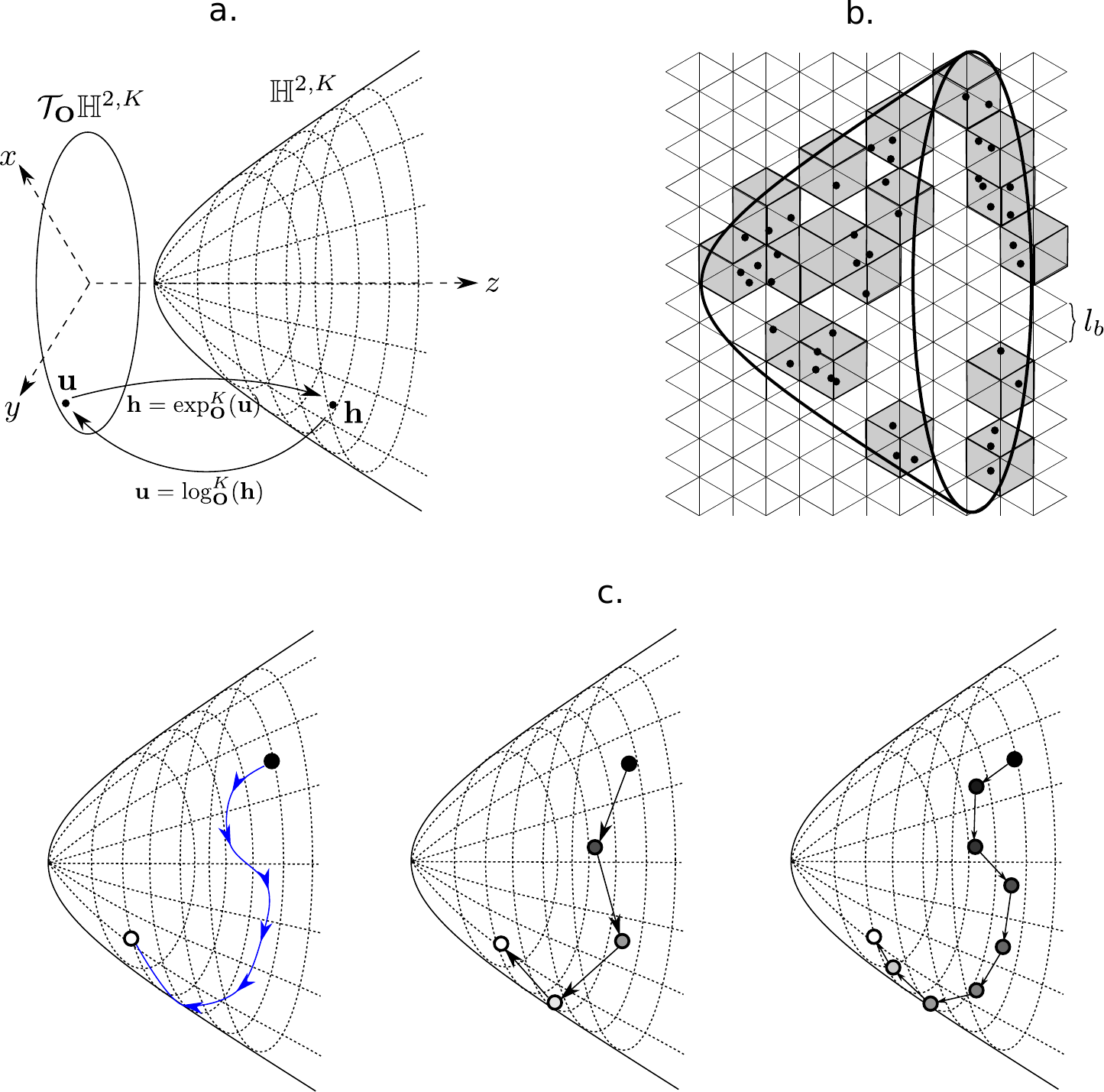}
    \caption{{\bf Point-particle trajectories on the hyperbolic manifold.} {\bf a.} The Lorentzian model for representing the 2-D Hyperbolic space $\mathbb{H}^{2,K}$ with constant negative curvature $-\frac{1}{K}$ as a hyperboloid; any point $\mathbf{u}$ on the tangent vector space at the origin, $\mathcal{T}_{\mathbf{O}}\mathbb{H}^{2,K}$ can be mapped to $\mathbb{H}^{2,K}$ via the exponential map (and back to the tangent space using the logarithmic map). {\bf b.} Box-counting of point-set over the hyperbolic plane. In this example, a 2-d hyperbolic set is covered using 17 3-d grids of size $l_b$. {\bf c.} Temporal scale transformation over the hyperboloid model; the blue curve on the leftmost figure represents the actual trajectory of a point-particle over the hyperboloid in time $[0, 1]$; middle and rightmost figures show the discrete time-scale transformation of the trajectory with $\Delta t=0.25$ and $0.125$, respectively.}
    \label{fig:hyperbolic-method}
\end{figure*}

\subsection{Scale transformation properties of real interaction networks}

Equipped with spatio-temporal scale transformation defined in Equation~\ref{eq:temporal-scale-transform}, we analyze six different temporal interaction networks arising from various real-world interactions -- Enron email interactions~\cite{enron-mail} ({\tt ia-email}), Reddit hyperlink interactions~\cite{reddit} ({\tt reddit-hyperlink}), Arxiv high-energy physics papers citations~\cite{nr} ({\tt ca-cit}), protein-protein interactions~\cite{DPPIN} ({\tt DPPIN-Babu}), Wikipedia talkpage interactions~\cite{superuser} ({\tt wiki-talk}) and user interactions via posts and comments on the Superuser forum~\cite{superuser} ({\tt superuser}) (see Table~\ref{tab:real-networks} for various statistics of these datasets). The choice of networks encompasses a wide variety of interactions: from user interactions over social platforms to protein-protein interactions. Furthermore, the list includes very slow-growing (e.g., citation networks) to very fast-growing (e.g., Reddit or Superuser) networks. We need to set different sizes for the time window $\Delta t$ in different networks to cope with this variation of interaction density.

In Figure~\ref{fig:real-networks}, we summarize the scale transformation properties of these networks. Equation~\ref{eq:temporal-fractal-dim} indicates that there can be three possible patterns for a network to exhibit: i) a linear variation of $\log(N_v/N_0)$ vs $\log(v)$ denoting scale-invariance, ii) a linear variation of $\log(N_v/N_0)$ with $\log(v)$ for larger values of $v$, denoting a finite fractal dimension but not scale-invariance, and iii) no linear relationship between $\log(N_v/N_0)$ and $\log(v)$, denoting a non-fractal network. The box-counting estimation of fractality and scale invariance is inherently statistical. Since the networks in consideration are finite, a small degree of noise can skew the presence (or absence) of linear relationships among observed samples. To deal with this, we fit two regression curves, one assuming $\log(N_v/N_0) \propto \log(v)$ (i.e., a finite fractal dimension) and another assuming $\log(N_v/N_0) \propto v$ (i.e., $d_v\to \infty$ as $v\to \infty$). As Figure~\ref{fig:real-networks} suggests, all the networks except {\tt DPPIN-Babu} and {\tt superuser} have a finite fractal dimension in the limiting case. We can calculate the value of the corresponding fractal dimensions from the slope of the regression lines -- $1.96$ for {\tt ia-email}, $3.76$ for {\tt reddit-hyperlink}, $2.45$ for {\tt ca-cit}, and $1.74$ for {\tt wiki-talk}. Among these fractal networks, only {\tt ia-email} shows scale-invariant characteristics under flow-scale transformations.

We further experiment on the scaling properties of some of these networks under purely spatial or temporal scale transformations (see {\em Supplementary}, Figures~S1 to S4). As we can observe, a finite fractal dimension under flow-scale transformation necessitates scale-invariance under both length- and time-scale transformations. Failure in either of these conditions would lead to non-fractal behaviour (like {\tt DPPIN-Babu}).

\subsection{Point-particle motion on hyperbolic surface}

One can intuitively identify the proposed flow-scale boxes as a potential {\em volume of influence} of a cluster of nodes over both space as well as time. As we go from smaller to larger time-scales as well length-scales, the chances of a node interacting with other nodes increase. This eludes one to draw an analogy between a temporal interaction network and a system of point particles moving randomly in space; the chance of two particles interacting with each other increases as they come closer in space. Such an analogy naturally presumes a latent geometry of the point-particle system.
It has been shown previously that an assumption of underlying non-Euclidean space explains the emergence of self-similarity in complex networks~\cite{self-similarity-metric-space-1, boguna2021network}. Following the precedence of linking hyperbolic geometry to the evolution of complex networks~\cite{hyperbolic-complex-network-1, hyperbolic-complex-network-2}, we investigate the spatial and temporal scale transformation properties of discrete objects on hyperbolic space to seek parallels with the temporal information networks. Please see {\em \em Methods} for constructing necessary structures of hyperbolic geometry in terms of the Lorentzian model.

By definition of hyperbolic geometry, points closer to the origin have a shorter geodesic distance compared to points further away (Figure~\ref{fig:hyperbolic-method} b.). This provides us with a natural analogy to complex networks. For example, a hub, which is reachable from other nodes with a shorter path, can be represented as a point near the origin (and peripheral nodes can be mapped further away). Furthermore, the Lorentzian model equips us with the property that one can define operations on the Euclidean tangent space with minimal complexity and map them back to the hyperbolic space using the exponential map. With this, we proceed to define the point-particle motion over $\mathbb{H}^{n,K}$. 

We define the process over the timespan $T:= [0,1]$. A point-particle is characterized by its position and velocity vectors on the tangent space at the origin, $\mathcal{T}_\mathbf{O}\mathbb{H}^{n,K}$. At any timestamp $t\in T$, a point-particle is initialized at position $\mathbf{u}(t)$ randomly over $\mathcal{T}_\mathbf{O}\mathbb{H}^{n,K}$. The `true' or latent trajectory of the particle is governed by a parametric estimation of the velocity given by
\begin{equation}
    \mathbf{\Tilde{v}}(t) = \sum_{i=0}^p\mathbf{v}_it^i
\end{equation}
where $\mathbf{v}_i$ are random parameters that characterize $\mathbf{\Tilde{v}}(t)$. This latent trajectory is intrinsic to the particle-system and, therefore, can not be observed directly. Instead, the system is realized in discrete timestamps $t_i\in T$, as shown in Figure~\ref{fig:hyperbolic-method} c. The position of a particle over $\mathcal{T}_\mathbf{O}\mathbb{H}^{n,K}$ at time $t_i$ can be calculated as
\begin{equation}
\label{eq:hyperbolic-motion}
    \mathbf{u}(t_i) = \mathbf{u}(t_{i-1}) + \mathbf{\Tilde{v}}(t_{i-1})(t_i-t_{i-1})
\end{equation}
and then mapped to $\mathbb{H}^{n,K}$ using the exponential map.
With equally spaced timestamps $t_i$'s and $\Delta t=t_{i}-t_{i-1}$, the sequence of particle positions approach the true trajectory as $\Delta t\to 0$. This $\Delta t$ serves the purpose of the time-scale transformation similar to Equation~\ref{eq:temporal-scale-transform}. 
Since in the hyperboloid model, $\mathbb{H}^{n,K}$ is immersed in $\mathbb{R}^{n+1}$, the natural choice of defining the length-scale transformation is via covering the particle system with $n+1$-dimensional grids of size $l_b$, as shown in Figure~\ref{fig:hyperbolic-method} b. With these two definitions, we can directly define the fractal dimension of this particle system as the same as in Equation~\ref{eq:temporal-fractal-dim}.

\begin{figure*}[!ht]
    \centering
    \includegraphics[width=\textwidth]{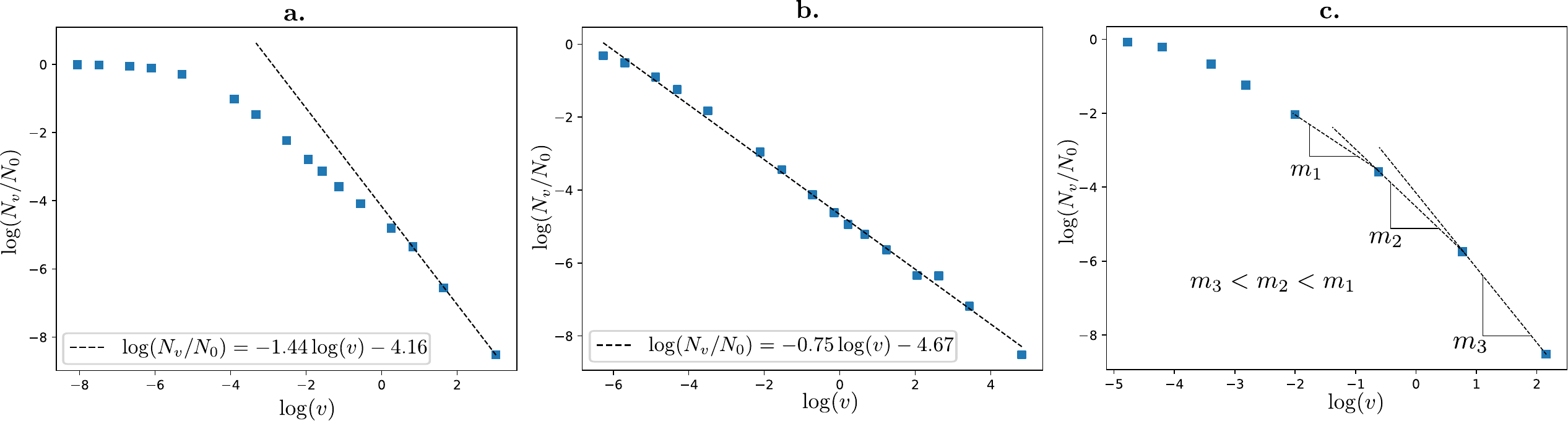}
    \caption{{\bf Scale transformation characteristics of point-particle motion.} {\bf a.} This plot corresponds to a 3-dimensional space of constant curvature $-1$, where the particle positions are sampled uniformly over the tangent space; though not scale-invariant, it has a finite fractal dimension at the limiting case. {\bf b.} This plot corresponds to a 3-dimensional space with the negative curvature increasing exponentially with time, and particles are sampled uniformly; it shows a scale-invariant characteristic against the transformation of $v$.  {\bf c.} This plot corresponds to a flat 3-dimensional space with the particle positions sampled from a Gaussian; evidently, the fractal dimension is not bounded, and therefore, the said structure does not possess fractal properties. }
    \label{fig:hyperbolic-results}
\end{figure*}

\subsection{Scale-invariance and time-varying curvature}

The point-particle motion described in the previous section can be majorly characterized by two parameters: the curvature of the space and the choice of the distribution from which the initial particle positions are sampled. 

We experiment with three different scenarios in terms of curvature. Our procedures defined in the previous section already demonstrate the constant negative curvature scenario: the particle position at any time $t$ is found via mapping $\mathbf{u}(t)$ to $\mathbb{H}^{n,K}$ using the exponential map (Equation~\ref{eq:exp-map}). An analogous phenomenon over a zero curvature (i.e., Euclidean) space can be achieved if $\mathbf{u}(t)$ and $\mathbf{\Tilde{v}}(t)$ are defined over $\mathbb{R}^{n+1}$ instead of $\mathcal{T}_\mathbf{O}\mathbb{H}^{n,K}$ and scale transformations are applied over them directly instead of applying an exponential map. Finally, we experiment with a space with negative curvature varying with time. This can be achieved using the following definition of the exponential map:
\begin{eqnarray}
    \label{eq:time-vary-curvature}
    \mathbf{h}(t_i) = \operatorname{exp}^{K(t_i)}_{\mathbf{O}(t_i)}(\mathbf{u}(t_i)) = \operatorname{cosh}\left (\frac{|\mathbf{u}(t_i)|_\mathcal{L}}{\sqrt{K(t_i)}}\right)\mathbf{O}(t_i) \nonumber \\
     + \sqrt{K(t_i)}\operatorname{sinh}\left (\frac{|\mathbf{u}(t_i)|_\mathcal{L}}{\sqrt{K(t_i)}} \right )\frac{\mathbf{u}(t_i)}{|\mathbf{u}(t_i)|_\mathcal{L}}
\end{eqnarray}
Here, the origin $\mathbf{O}$ shifts with time since it is defined explicitly using the curvature of the space. 

Additionally, we experiment with the particle positions initially sampled from a Gaussian vs a uniform distribution, both centred at the origin of $\mathbb{R}^{n+1}$. We find that the three distinctive patterns of scale transformation that we observed in the case of the real networks emerge with specific choices of curvature and distributions in the point-particle motion as well, as shown in Figure~\ref{fig:hyperbolic-results}. When the particle positions are sampled uniformly over a constant curvature space, we observe that the fractal dimension, albeit non-constant, converges to a finite value as $v\to \infty$. We conclude that {\em a constant negative curvature space almost always guarantees a finite fractal dimension but not a scale-invariant one} (Figure~\ref{fig:hyperbolic-results} a. shows the variation with $K=1$ and $n=3$). Scale-invariance exclusively arises when we move to space with time-varying negative curvature, specifically, $-\frac{1}{K_0}e^{at}$ for some positive constants $K_0$ and $a$; in the example shown in Figure~\ref{fig:hyperbolic-results} b., $K_0=1$, $a=1$, $n=3$. Finally, a particle set initialized from a Gaussian over the flat Euclidean space fails to exhibit a finite fractal dimension (Figure~\ref{fig:hyperbolic-results} c. with $n=3$ and a zero mean unit standard deviation Gaussian). 

\section{Discussion}

We developed a method for spatio-temporal scale transformations over temporal interaction networks lacking predefined geometry. Our described method relied upon a coarsening of time-scale to build static snapshots of the interaction stream that can be treated as networks induced by simultaneous interactions. Our analysis of multiple real-world interaction networks suggested the emergence of finite fractal dimensions under our proposed notion of flow-scale transformations. In an attempt to link such invariance properties to a latent geometry of the networks, we simulated point-particle trajectories over hyperbolic space. These point-particle systems served as a parallel to temporal interaction networks, similar to the previous notion of random geometric graphs in which the probability of forming an edge between any two nodes embedded in a geometry is proportional to the distance between these nodes. Emulating the time-scale coarsening and grid-covering over the ambient space of these particle systems revealed that a scale-invariant system is possible only when the underlying hyperbolic space has a time-varying curvature (to be specific, an exponentially increasing value of negative curvature). Flat geometries fail to show a bounded fractal dimension.

One major challenge in discussing the scaling properties of real-world networks is that these networks are finite-size, and scale-invariant properties are often hidden under noises~\cite{finite-size-issue-1, finite-size-issues-2}. Our work also exhibits the same limitations -- most of the temporal interaction networks we considered are of moderate size, and their scaling characteristics are skewed when the flow-scale boxes are small. However, the strong parallel between the scaling properties of the real-world networks and those of the point-particle trajectories in latent geometric spaces materializes the phenomena beyond domain-dependent artefacts. Our findings are particularly useful towards modeling social/biological networks. Linking the latent space model with the scaling properties of temporal interaction networks, our findings can broaden the understanding of network embedding methods --- a core problem in the predictive modeling of interaction networks. Finally, our work may facilitate the applications of more general understandings of critical processes in statistical physics~\cite{other-scale-1, other-scale-2, other-scale-3, other-scales-4} that depict scale-invariance to the specific problems of network science.

\section{Materials and Methods}

\subsection{Defining flow-scale boxes}

The definition of flow-scale boxes requires simultaneous scale transformation over space as well as time. To define the time-scale transformation, we use successive coarsening of the temporal measure that instantiates the interaction timestamps in the following manner.

Let a temporal interaction network be defined as a chronologically ordered sequence of time-stamped interactions $I(t_0, t_N) = \{(v_i, v_j, t_k)\}$, where $t_0$ and $t_N$ are the start and end times of the interactions, respectively, and $v_i$ and $v_j$ are any two nodes interacting at time $t_k$. The discrete nature of these timestamps presumes that the observation was done at a certain degree of coarsening, i.e., they are measured to the nearest integer value on a specific unit of time like seconds, hours, or days. To illustrate further with an example, if the timestamps are reported in integer seconds, and if an interaction actually occurred at time $3.05$ seconds, then the timestamp that will be associated with the said interaction will be $3$ seconds. Therefore, it will be treated as contemporary with another interaction happening at a time of $3.14$ seconds. In this example, the chosen scale of time is $1$ second. We extend this to any arbitrary time-scale to renormalize the network. Given a time-scale $\Delta t$, the renormalized interaction network can be defined as:
\begin{equation}
    I_{\Delta t}(t_0, t_N) = \{(v_i, v_j, t_0+m\Delta t)|m\in \{0, \cdots, \lfloor \frac{N}{\Delta t}\rfloor\}
\end{equation}
for any $(v_i, v_j, t_k)\in I(t_0, t_N)$ such that $t_k\in [t_0+m\Delta t, t_0+(m+1)\Delta t)$. Such a time-scale renormalization treats any interaction happening within the $\Delta t$ window as contemporary, thereby segmenting $I(t_0, t_N)$ into $\lfloor \frac{N}{\Delta t}\rfloor$ static snapshots. Each of these snapshots can be considered a static network constructed by nodes interacting with each other simultaneously (under the time-scale renormalization). The $m$-th snapshot is then defined as $G_m = \{V_m, E_m\}$ with $V_m$ and $E_m$ being the set of nodes and edges, respectively such that, for any $e=(v_i,v_j)\in E_m$, there exists at least one interaction $(v_i, v_j, t_0+m\Delta t)\in I_{\Delta t}(t_0, t_N)$. 

These snapshots now bear a geometric structure (i.e., the distance between two nodes can be defined based on the path length) conditioned upon the concept of simultaneity introduced by the coarsening of the time-scale. This allows us to apply box-counting to initiate length-scale renormalization.

Next, we apply box-covering to perform length-scale renormalization on the snapshots. We use the Maximum Excluded Mass Burning (MEMB) algorithm~\cite{MEMB} to count the number of boxes of a given size needed to cover a given snapshot $G_m$. MEMB defines boxes with their radius instead of diameter. A box of radius $r_b$ is a connected subgraph with a root node of arbitrary choice such that any node in the subgraph is at most $r_B$ distance away from the root node. The number of boxes to cover the interaction network is then,
\begin{equation}
    N_v = \sum_m \operatorname{MEMB}(G_m)
\end{equation}
where $N_v$ is the number of flow-scale boxes of time-scale $\Delta t$ and length-scale $l_b=2r_b$ needed to cover the temporal interaction network $I(t_0, t_N)$. Details of the choices of parameter sizes for different datasets are described in {\em Supplementary}, Table S1.

\subsection{Lorentzian model of hyperbolic geometry}

We start with the $n$-dimensional, Lorentzian Hyperboloid model of the hyperbolic space, denoted by $\mathbb{H}^{n,K}$ of constant negative curvature $-\frac{1}{K}$, embedded in an $(n+1)$-dimensional space (i.e., $\mathbb{H}^{n,K}\subset \mathbb{R}^{n+1}$). For any two points $\mathbf{h},\mathbf{g}\in \mathbb{R}^{n+1}$, we have the Minkowski inner product:
\begin{equation}
    \langle \mathbf{h},\mathbf{g} \rangle_\mathcal{L} = -h_0g_0 + \sum_{i=1}^n h_ig_i
\end{equation}
For any $\mathbf{h}\in \mathbb{H}^{n,K}$, we have $\langle \mathbf{h}, \mathbf{h}\rangle_{\mathcal{L}} = -K$. The tangent space to $\mathbb{H}^{n,K}$ at point $\mathbf{x}\in \mathbb{H}^{n,K}$ is defined as, 
\begin{equation}
    \mathcal{T}_\mathbf{x}\mathbb{H}^{2,K}=\{\mathbf{v}|\langle \mathbf{v}, \mathbf{x}\rangle_{\mathcal{L}}=0\}
\end{equation}
For any point $\mathbf{u}\in \mathcal{T}_\mathbf{x}\mathbb{H}^{n,K}$, the Minkowski norm is defined by $|\mathbf{u}|_\mathcal{L} = \langle \mathbf{u}, \mathbf{u}\rangle_{\mathcal{L}}$. The origin of $\mathbb{H}^{n,K}$ is defined as $\mathbf{O}$ where $O_0=\sqrt{K}$ and $O_i=0$ for all $i\neq 0$. One can define a bijective map between $\mathbb{H}^{n,K}$ and $\mathcal{T}_\mathbf{x}\mathbb{H}^{n,K}$, also called the {\em exponential map}, as,
\begin{equation}
\label{eq:exp-map}
    \mathbf{h} = \operatorname{exp}^{K}_\mathbf{x}(\mathbf{u}) = \operatorname{cosh}\left (\frac{|\mathbf{u}|_\mathcal{L}}{\sqrt{K}}\right)\mathbf{x} + \sqrt{K}\operatorname{sinh}\left (\frac{|\mathbf{u}|_\mathcal{L}}{\sqrt{K}} \right )\frac{\mathbf{u}}{|\mathbf{u}|_\mathcal{L}}
\end{equation}
for $\mathbf{h},\mathbf{x}\in \mathbb{H}^{n,K}$ and $\mathbf{u}\in \mathcal{T}_\mathbf{x}\mathbb{H}^{n,K}$. The inverse of the exponential map is called the logarithmic map and is defined as,
\begin{equation}
    \mathbf{u} = \operatorname{log}^K_\mathbf{x}(\mathbf{h})= d^K_\mathcal{L}(\mathbf{x},\mathbf{h})\frac{\mathbf{h} + \frac{1}{K}\langle \mathbf{h}, \mathbf{x}\rangle_\mathcal{L}\mathbf{x}}{|\mathbf{h} + \frac{1}{K}\langle \mathbf{h}, \mathbf{x}\rangle_\mathcal{L}\mathbf{x}|_\mathcal{L}}
\end{equation}
where $d^K_\mathcal{L}(\mathbf{x},\mathbf{h})=\sqrt{K}\operatorname{arccosh}(-\langle \mathbf{x},\mathbf{h} \rangle_\mathcal{L}/K)$ is the distance (length of the geodesic curve) between $\mathbf{x}$ and $\mathbf{h}$.

\section*{Data and Source Code Availability}
The datasets used in this paper are publicly available at \href{https://snap.stanford.edu}{https://snap.stanford.edu} and \href{https://networkrepository.com}{https://networkrepository.com}. The source codes is available at \href{https://github.com/Subha0009/Temporal-network-self-similarity}{https://github.com/Subha0009/Temporal-network-self-similarity}.

\section*{Supplementary Information}
The supplementary contains additional information related to network datasets, experimental setup, and results.

\section*{Acknowledgement}T.C. would like to acknowledge the support of the Ramanujan Fellowship (SB/S2/RJN-073/2017), funded by the Science and Engineering Research Board (SERB), India.


%


\begin{thebibliography}{42}%
\makeatletter
\providecommand \@ifxundefined [1]{%
 \@ifx{#1\undefined}
}%
\providecommand \@ifnum [1]{%
 \ifnum #1\expandafter \@firstoftwo
 \else \expandafter \@secondoftwo
 \fi
}%
\providecommand \@ifx [1]{%
 \ifx #1\expandafter \@firstoftwo
 \else \expandafter \@secondoftwo
 \fi
}%
\providecommand \natexlab [1]{#1}%
\providecommand \enquote  [1]{``#1''}%
\providecommand \bibnamefont  [1]{#1}%
\providecommand \bibfnamefont [1]{#1}%
\providecommand \citenamefont [1]{#1}%
\providecommand \href@noop [0]{\@secondoftwo}%
\providecommand \href [0]{\begingroup \@sanitize@url \@href}%
\providecommand \@href[1]{\@@startlink{#1}\@@href}%
\providecommand \@@href[1]{\endgroup#1\@@endlink}%
\providecommand \@sanitize@url [0]{\catcode `\\12\catcode `\$12\catcode `\&12\catcode `\#12\catcode `\^12\catcode `\_12\catcode `\%12\relax}%
\providecommand \@@startlink[1]{}%
\providecommand \@@endlink[0]{}%
\providecommand \url  [0]{\begingroup\@sanitize@url \@url }%
\providecommand \@url [1]{\endgroup\@href {#1}{\urlprefix }}%
\providecommand \urlprefix  [0]{URL }%
\providecommand \Eprint [0]{\href }%
\providecommand \doibase [0]{https://doi.org/}%
\providecommand \selectlanguage [0]{\@gobble}%
\providecommand \bibinfo  [0]{\@secondoftwo}%
\providecommand \bibfield  [0]{\@secondoftwo}%
\providecommand \translation [1]{[#1]}%
\providecommand \BibitemOpen [0]{}%
\providecommand \bibitemStop [0]{}%
\providecommand \bibitemNoStop [0]{.\EOS\space}%
\providecommand \EOS [0]{\spacefactor3000\relax}%
\providecommand \BibitemShut  [1]{\csname bibitem#1\endcsname}%
\let\auto@bib@innerbib\@empty
\bibitem [{\citenamefont {Paxson}\ and\ \citenamefont {Varanasi}(2013)}]{general-self-similarity-1}%
  \BibitemOpen
  \bibfield  {author} {\bibinfo {author} {\bibfnamefont {A.~T.}\ \bibnamefont {Paxson}}\ and\ \bibinfo {author} {\bibfnamefont {K.~K.}\ \bibnamefont {Varanasi}},\ }\bibfield  {title} {\bibinfo {title} {Self-similarity of contact line depinning from textured surfaces},\ }\href {https://doi.org/10.1038/ncomms2482} {\bibfield  {journal} {\bibinfo  {journal} {Nature Communications}\ }\textbf {\bibinfo {volume} {4}},\ \bibinfo {pages} {1492} (\bibinfo {year} {2013})}\BibitemShut {NoStop}%
\bibitem [{\citenamefont {Rydelek}\ and\ \citenamefont {Sacks}(1989)}]{general-self-similarity-2}%
  \BibitemOpen
  \bibfield  {author} {\bibinfo {author} {\bibfnamefont {P.~A.}\ \bibnamefont {Rydelek}}\ and\ \bibinfo {author} {\bibfnamefont {I.~S.}\ \bibnamefont {Sacks}},\ }\bibfield  {title} {\bibinfo {title} {Testing the completeness of earthquake catalogues and the hypothesis of self-similarity},\ }\href {https://doi.org/10.1038/337251a0} {\bibfield  {journal} {\bibinfo  {journal} {Nature}\ }\textbf {\bibinfo {volume} {337}},\ \bibinfo {pages} {251} (\bibinfo {year} {1989})}\BibitemShut {NoStop}%
\bibitem [{\citenamefont {Hs{\"u}}\ and\ \citenamefont {Hs{\"u}}(1991)}]{temporal-scale-2}%
  \BibitemOpen
  \bibfield  {author} {\bibinfo {author} {\bibfnamefont {K.~J.}\ \bibnamefont {Hs{\"u}}}\ and\ \bibinfo {author} {\bibfnamefont {A.}~\bibnamefont {Hs{\"u}}},\ }\bibfield  {title} {\bibinfo {title} {Self-similarity of the" 1/f noise" called music.},\ }\href@noop {} {\bibfield  {journal} {\bibinfo  {journal} {Proceedings of the National Academy of Sciences}\ }\textbf {\bibinfo {volume} {88}},\ \bibinfo {pages} {3507} (\bibinfo {year} {1991})}\BibitemShut {NoStop}%
\bibitem [{\citenamefont {Simini}\ \emph {et~al.}(2010)\citenamefont {Simini}, \citenamefont {Anfodillo}, \citenamefont {Carrer}, \citenamefont {Banavar},\ and\ \citenamefont {Maritan}}]{general-self-similarity-4}%
  \BibitemOpen
  \bibfield  {author} {\bibinfo {author} {\bibfnamefont {F.}~\bibnamefont {Simini}}, \bibinfo {author} {\bibfnamefont {T.}~\bibnamefont {Anfodillo}}, \bibinfo {author} {\bibfnamefont {M.}~\bibnamefont {Carrer}}, \bibinfo {author} {\bibfnamefont {J.~R.}\ \bibnamefont {Banavar}},\ and\ \bibinfo {author} {\bibfnamefont {A.}~\bibnamefont {Maritan}},\ }\bibfield  {title} {\bibinfo {title} {Self-similarity and scaling in forest communities},\ }\href@noop {} {\bibfield  {journal} {\bibinfo  {journal} {Proceedings of the National Academy of Sciences}\ }\textbf {\bibinfo {volume} {107}},\ \bibinfo {pages} {7658} (\bibinfo {year} {2010})}\BibitemShut {NoStop}%
\bibitem [{\citenamefont {Teo}\ and\ \citenamefont {Zhang}(1991)}]{general-self-similarity-5}%
  \BibitemOpen
  \bibfield  {author} {\bibinfo {author} {\bibfnamefont {B.~K.}\ \bibnamefont {Teo}}\ and\ \bibinfo {author} {\bibfnamefont {H.}~\bibnamefont {Zhang}},\ }\bibfield  {title} {\bibinfo {title} {Clusters of clusters: self-organization and self-similarity in the intermediate stages of cluster growth of au-ag supraclusters.},\ }\href@noop {} {\bibfield  {journal} {\bibinfo  {journal} {Proceedings of the National Academy of Sciences}\ }\textbf {\bibinfo {volume} {88}},\ \bibinfo {pages} {5067} (\bibinfo {year} {1991})}\BibitemShut {NoStop}%
\bibitem [{\citenamefont {Song}\ \emph {et~al.}(2005)\citenamefont {Song}, \citenamefont {Havlin},\ and\ \citenamefont {Makse}}]{self-similar-2005}%
  \BibitemOpen
  \bibfield  {author} {\bibinfo {author} {\bibfnamefont {C.}~\bibnamefont {Song}}, \bibinfo {author} {\bibfnamefont {S.}~\bibnamefont {Havlin}},\ and\ \bibinfo {author} {\bibfnamefont {H.~A.}\ \bibnamefont {Makse}},\ }\bibfield  {title} {\bibinfo {title} {Self-similarity of complex networks},\ }\href@noop {} {\bibfield  {journal} {\bibinfo  {journal} {Nature}\ }\textbf {\bibinfo {volume} {433}},\ \bibinfo {pages} {392} (\bibinfo {year} {2005})}\BibitemShut {NoStop}%
\bibitem [{\citenamefont {Gallos}\ \emph {et~al.}(2007{\natexlab{a}})\citenamefont {Gallos}, \citenamefont {Song},\ and\ \citenamefont {Makse}}]{box-cover-1}%
  \BibitemOpen
  \bibfield  {author} {\bibinfo {author} {\bibfnamefont {L.~K.}\ \bibnamefont {Gallos}}, \bibinfo {author} {\bibfnamefont {C.}~\bibnamefont {Song}},\ and\ \bibinfo {author} {\bibfnamefont {H.~A.}\ \bibnamefont {Makse}},\ }\bibfield  {title} {\bibinfo {title} {A review of fractality and self-similarity in complex networks},\ }\href@noop {} {\bibfield  {journal} {\bibinfo  {journal} {Physica A: Statistical Mechanics and its Applications}\ }\textbf {\bibinfo {volume} {386}},\ \bibinfo {pages} {686} (\bibinfo {year} {2007}{\natexlab{a}})}\BibitemShut {NoStop}%
\bibitem [{\citenamefont {Zheng}\ \emph {et~al.}(2020)\citenamefont {Zheng}, \citenamefont {Allard}, \citenamefont {Hagmann}, \citenamefont {Alem{\'a}n-G{\'o}mez},\ and\ \citenamefont {Serrano}}]{general-self-similarity-3}%
  \BibitemOpen
  \bibfield  {author} {\bibinfo {author} {\bibfnamefont {M.}~\bibnamefont {Zheng}}, \bibinfo {author} {\bibfnamefont {A.}~\bibnamefont {Allard}}, \bibinfo {author} {\bibfnamefont {P.}~\bibnamefont {Hagmann}}, \bibinfo {author} {\bibfnamefont {Y.}~\bibnamefont {Alem{\'a}n-G{\'o}mez}},\ and\ \bibinfo {author} {\bibfnamefont {M.~{\'A}.}\ \bibnamefont {Serrano}},\ }\bibfield  {title} {\bibinfo {title} {Geometric renormalization unravels self-similarity of the multiscale human connectome},\ }\href@noop {} {\bibfield  {journal} {\bibinfo  {journal} {Proceedings of the National Academy of Sciences}\ }\textbf {\bibinfo {volume} {117}},\ \bibinfo {pages} {20244} (\bibinfo {year} {2020})}\BibitemShut {NoStop}%
\bibitem [{\citenamefont {Gallos}\ \emph {et~al.}(2007{\natexlab{b}})\citenamefont {Gallos}, \citenamefont {Song},\ and\ \citenamefont {Makse}}]{gallos2007review}%
  \BibitemOpen
  \bibfield  {author} {\bibinfo {author} {\bibfnamefont {L.~K.}\ \bibnamefont {Gallos}}, \bibinfo {author} {\bibfnamefont {C.}~\bibnamefont {Song}},\ and\ \bibinfo {author} {\bibfnamefont {H.~A.}\ \bibnamefont {Makse}},\ }\bibfield  {title} {\bibinfo {title} {A review of fractality and self-similarity in complex networks},\ }\href@noop {} {\bibfield  {journal} {\bibinfo  {journal} {Physica A: Statistical Mechanics and its Applications}\ }\textbf {\bibinfo {volume} {386}},\ \bibinfo {pages} {686} (\bibinfo {year} {2007}{\natexlab{b}})}\BibitemShut {NoStop}%
\bibitem [{\citenamefont {Proekt}\ \emph {et~al.}(2012)\citenamefont {Proekt}, \citenamefont {Banavar}, \citenamefont {Maritan},\ and\ \citenamefont {Pfaff}}]{temporal-scale-1}%
  \BibitemOpen
  \bibfield  {author} {\bibinfo {author} {\bibfnamefont {A.}~\bibnamefont {Proekt}}, \bibinfo {author} {\bibfnamefont {J.~R.}\ \bibnamefont {Banavar}}, \bibinfo {author} {\bibfnamefont {A.}~\bibnamefont {Maritan}},\ and\ \bibinfo {author} {\bibfnamefont {D.~W.}\ \bibnamefont {Pfaff}},\ }\bibfield  {title} {\bibinfo {title} {Scale invariance in the dynamics of spontaneous behavior},\ }\href@noop {} {\bibfield  {journal} {\bibinfo  {journal} {Proceedings of the National Academy of Sciences}\ }\textbf {\bibinfo {volume} {109}},\ \bibinfo {pages} {10564} (\bibinfo {year} {2012})}\BibitemShut {NoStop}%
\bibitem [{\citenamefont {Krug}(1997)}]{scale-invariance-seminal-1}%
  \BibitemOpen
  \bibfield  {author} {\bibinfo {author} {\bibfnamefont {J.}~\bibnamefont {Krug}},\ }\bibfield  {title} {\bibinfo {title} {Origins of scale invariance in growth processes},\ }\href@noop {} {\bibfield  {journal} {\bibinfo  {journal} {Advances in Physics}\ }\textbf {\bibinfo {volume} {46}},\ \bibinfo {pages} {139} (\bibinfo {year} {1997})}\BibitemShut {NoStop}%
\bibitem [{\citenamefont {Sornette}(1998{\natexlab{a}})}]{scale-invariance-seminal-2}%
  \BibitemOpen
  \bibfield  {author} {\bibinfo {author} {\bibfnamefont {D.}~\bibnamefont {Sornette}},\ }\bibfield  {title} {\bibinfo {title} {Discrete-scale invariance and complex dimensions},\ }\href@noop {} {\bibfield  {journal} {\bibinfo  {journal} {Physics reports}\ }\textbf {\bibinfo {volume} {297}},\ \bibinfo {pages} {239} (\bibinfo {year} {1998}{\natexlab{a}})}\BibitemShut {NoStop}%
\bibitem [{\citenamefont {Cohen}\ and\ \citenamefont {Havlin}(2004)}]{cohen2004fractal}%
  \BibitemOpen
  \bibfield  {author} {\bibinfo {author} {\bibfnamefont {R.}~\bibnamefont {Cohen}}\ and\ \bibinfo {author} {\bibfnamefont {S.}~\bibnamefont {Havlin}},\ }\bibfield  {title} {\bibinfo {title} {Fractal dimensions of percolating networks},\ }\href@noop {} {\bibfield  {journal} {\bibinfo  {journal} {Physica A: Statistical Mechanics and its Applications}\ }\textbf {\bibinfo {volume} {336}},\ \bibinfo {pages} {6} (\bibinfo {year} {2004})}\BibitemShut {NoStop}%
\bibitem [{\citenamefont {Stanley}\ \emph {et~al.}(2000)\citenamefont {Stanley}, \citenamefont {Amaral}, \citenamefont {Gopikrishnan}, \citenamefont {Ivanov}, \citenamefont {Keitt},\ and\ \citenamefont {Plerou}}]{scale-invariance-seminal-3}%
  \BibitemOpen
  \bibfield  {author} {\bibinfo {author} {\bibfnamefont {H.~E.}\ \bibnamefont {Stanley}}, \bibinfo {author} {\bibfnamefont {L.}~\bibnamefont {Amaral}}, \bibinfo {author} {\bibfnamefont {P.}~\bibnamefont {Gopikrishnan}}, \bibinfo {author} {\bibfnamefont {P.~C.}\ \bibnamefont {Ivanov}}, \bibinfo {author} {\bibfnamefont {T.}~\bibnamefont {Keitt}},\ and\ \bibinfo {author} {\bibfnamefont {V.}~\bibnamefont {Plerou}},\ }\bibfield  {title} {\bibinfo {title} {Scale invariance and universality: organizing principles in complex systems},\ }\href@noop {} {\bibfield  {journal} {\bibinfo  {journal} {Physica A: Statistical Mechanics and its Applications}\ }\textbf {\bibinfo {volume} {281}},\ \bibinfo {pages} {60} (\bibinfo {year} {2000})}\BibitemShut {NoStop}%
\bibitem [{\citenamefont {Guimera}\ \emph {et~al.}(2003)\citenamefont {Guimera}, \citenamefont {Danon}, \citenamefont {Diaz-Guilera}, \citenamefont {Giralt},\ and\ \citenamefont {Arenas}}]{guimera2003self}%
  \BibitemOpen
  \bibfield  {author} {\bibinfo {author} {\bibfnamefont {R.}~\bibnamefont {Guimera}}, \bibinfo {author} {\bibfnamefont {L.}~\bibnamefont {Danon}}, \bibinfo {author} {\bibfnamefont {A.}~\bibnamefont {Diaz-Guilera}}, \bibinfo {author} {\bibfnamefont {F.}~\bibnamefont {Giralt}},\ and\ \bibinfo {author} {\bibfnamefont {A.}~\bibnamefont {Arenas}},\ }\bibfield  {title} {\bibinfo {title} {Self-similar community structure in a network of human interactions},\ }\href@noop {} {\bibfield  {journal} {\bibinfo  {journal} {Physical review E}\ }\textbf {\bibinfo {volume} {68}},\ \bibinfo {pages} {065103} (\bibinfo {year} {2003})}\BibitemShut {NoStop}%
\bibitem [{\citenamefont {Sornette}(1998{\natexlab{b}})}]{scale-invariance-definition}%
  \BibitemOpen
  \bibfield  {author} {\bibinfo {author} {\bibfnamefont {D.}~\bibnamefont {Sornette}},\ }\bibfield  {title} {\bibinfo {title} {Discrete-scale invariance and complex dimensions},\ }\href@noop {} {\bibfield  {journal} {\bibinfo  {journal} {Physics reports}\ }\textbf {\bibinfo {volume} {297}},\ \bibinfo {pages} {239} (\bibinfo {year} {1998}{\natexlab{b}})}\BibitemShut {NoStop}%
\bibitem [{\citenamefont {Song}\ \emph {et~al.}(2006)\citenamefont {Song}, \citenamefont {Havlin},\ and\ \citenamefont {Makse}}]{self-similar-complex-networks-2}%
  \BibitemOpen
  \bibfield  {author} {\bibinfo {author} {\bibfnamefont {C.}~\bibnamefont {Song}}, \bibinfo {author} {\bibfnamefont {S.}~\bibnamefont {Havlin}},\ and\ \bibinfo {author} {\bibfnamefont {H.~A.}\ \bibnamefont {Makse}},\ }\bibfield  {title} {\bibinfo {title} {Origins of fractality in the growth of complex networks},\ }\href@noop {} {\bibfield  {journal} {\bibinfo  {journal} {Nature physics}\ }\textbf {\bibinfo {volume} {2}},\ \bibinfo {pages} {275} (\bibinfo {year} {2006})}\BibitemShut {NoStop}%
\bibitem [{\citenamefont {Goh}\ \emph {et~al.}(2006)\citenamefont {Goh}, \citenamefont {Salvi}, \citenamefont {Kahng},\ and\ \citenamefont {Kim}}]{self-similar-complex-networks-3}%
  \BibitemOpen
  \bibfield  {author} {\bibinfo {author} {\bibfnamefont {K.-I.}\ \bibnamefont {Goh}}, \bibinfo {author} {\bibfnamefont {G.}~\bibnamefont {Salvi}}, \bibinfo {author} {\bibfnamefont {B.}~\bibnamefont {Kahng}},\ and\ \bibinfo {author} {\bibfnamefont {D.}~\bibnamefont {Kim}},\ }\bibfield  {title} {\bibinfo {title} {Skeleton and fractal scaling in complex networks},\ }\href@noop {} {\bibfield  {journal} {\bibinfo  {journal} {Physical review letters}\ }\textbf {\bibinfo {volume} {96}},\ \bibinfo {pages} {018701} (\bibinfo {year} {2006})}\BibitemShut {NoStop}%
\bibitem [{\citenamefont {Serrano}\ \emph {et~al.}(2008)\citenamefont {Serrano}, \citenamefont {Krioukov},\ and\ \citenamefont {Bogun{\'a}}}]{self-similarity-metric-space-1}%
  \BibitemOpen
  \bibfield  {author} {\bibinfo {author} {\bibfnamefont {M.~{\'A}.}\ \bibnamefont {Serrano}}, \bibinfo {author} {\bibfnamefont {D.}~\bibnamefont {Krioukov}},\ and\ \bibinfo {author} {\bibfnamefont {M.}~\bibnamefont {Bogun{\'a}}},\ }\bibfield  {title} {\bibinfo {title} {Self-similarity of complex networks and hidden metric spaces},\ }\href@noop {} {\bibfield  {journal} {\bibinfo  {journal} {Physical review letters}\ }\textbf {\bibinfo {volume} {100}},\ \bibinfo {pages} {078701} (\bibinfo {year} {2008})}\BibitemShut {NoStop}%
\bibitem [{\citenamefont {Kim}\ \emph {et~al.}(2007{\natexlab{a}})\citenamefont {Kim}, \citenamefont {Goh}, \citenamefont {Kahng},\ and\ \citenamefont {Kim}}]{kim2007fractality}%
  \BibitemOpen
  \bibfield  {author} {\bibinfo {author} {\bibfnamefont {J.~S.}\ \bibnamefont {Kim}}, \bibinfo {author} {\bibfnamefont {K.-I.}\ \bibnamefont {Goh}}, \bibinfo {author} {\bibfnamefont {B.}~\bibnamefont {Kahng}},\ and\ \bibinfo {author} {\bibfnamefont {D.}~\bibnamefont {Kim}},\ }\bibfield  {title} {\bibinfo {title} {Fractality and self-similarity in scale-free networks},\ }\href@noop {} {\bibfield  {journal} {\bibinfo  {journal} {New Journal of Physics}\ }\textbf {\bibinfo {volume} {9}},\ \bibinfo {pages} {177} (\bibinfo {year} {2007}{\natexlab{a}})}\BibitemShut {NoStop}%
\bibitem [{\citenamefont {Tsybakov}\ and\ \citenamefont {Georganas}(1998)}]{tsybakov1998self}%
  \BibitemOpen
  \bibfield  {author} {\bibinfo {author} {\bibfnamefont {B.}~\bibnamefont {Tsybakov}}\ and\ \bibinfo {author} {\bibfnamefont {N.~D.}\ \bibnamefont {Georganas}},\ }\bibfield  {title} {\bibinfo {title} {Self-similar processes in communications networks},\ }\href@noop {} {\bibfield  {journal} {\bibinfo  {journal} {IEEE Transactions on Information Theory}\ }\textbf {\bibinfo {volume} {44}},\ \bibinfo {pages} {1713} (\bibinfo {year} {1998})}\BibitemShut {NoStop}%
\bibitem [{\citenamefont {Kim}\ \emph {et~al.}(2007{\natexlab{b}})\citenamefont {Kim}, \citenamefont {Goh}, \citenamefont {Kahng},\ and\ \citenamefont {Kim}}]{box-cover-2}%
  \BibitemOpen
  \bibfield  {author} {\bibinfo {author} {\bibfnamefont {J.~S.}\ \bibnamefont {Kim}}, \bibinfo {author} {\bibfnamefont {K.-I.}\ \bibnamefont {Goh}}, \bibinfo {author} {\bibfnamefont {B.}~\bibnamefont {Kahng}},\ and\ \bibinfo {author} {\bibfnamefont {D.}~\bibnamefont {Kim}},\ }\bibfield  {title} {\bibinfo {title} {Fractality and self-similarity in scale-free networks},\ }\href@noop {} {\bibfield  {journal} {\bibinfo  {journal} {New Journal of Physics}\ }\textbf {\bibinfo {volume} {9}},\ \bibinfo {pages} {177} (\bibinfo {year} {2007}{\natexlab{b}})}\BibitemShut {NoStop}%
\bibitem [{\citenamefont {Zhou}\ \emph {et~al.}(2007)\citenamefont {Zhou}, \citenamefont {Jiang},\ and\ \citenamefont {Sornette}}]{box-cover-3}%
  \BibitemOpen
  \bibfield  {author} {\bibinfo {author} {\bibfnamefont {W.-X.}\ \bibnamefont {Zhou}}, \bibinfo {author} {\bibfnamefont {Z.-Q.}\ \bibnamefont {Jiang}},\ and\ \bibinfo {author} {\bibfnamefont {D.}~\bibnamefont {Sornette}},\ }\bibfield  {title} {\bibinfo {title} {Exploring self-similarity of complex cellular networks: The edge-covering method with simulated annealing and log-periodic sampling},\ }\href@noop {} {\bibfield  {journal} {\bibinfo  {journal} {Physica A: Statistical Mechanics and its Applications}\ }\textbf {\bibinfo {volume} {375}},\ \bibinfo {pages} {741} (\bibinfo {year} {2007})}\BibitemShut {NoStop}%
\bibitem [{\citenamefont {Klimovskaia}\ \emph {et~al.}(2020)\citenamefont {Klimovskaia}, \citenamefont {Lopez-Paz}, \citenamefont {Bottou},\ and\ \citenamefont {Nickel}}]{Hyperbolic-graph-embeddings-1}%
  \BibitemOpen
  \bibfield  {author} {\bibinfo {author} {\bibfnamefont {A.}~\bibnamefont {Klimovskaia}}, \bibinfo {author} {\bibfnamefont {D.}~\bibnamefont {Lopez-Paz}}, \bibinfo {author} {\bibfnamefont {L.}~\bibnamefont {Bottou}},\ and\ \bibinfo {author} {\bibfnamefont {M.}~\bibnamefont {Nickel}},\ }\bibfield  {title} {\bibinfo {title} {Poincar{\'e} maps for analyzing complex hierarchies in single-cell data},\ }\href {https://doi.org/10.1038/s41467-020-16822-4} {\bibfield  {journal} {\bibinfo  {journal} {Nature Communications}\ }\textbf {\bibinfo {volume} {11}},\ \bibinfo {pages} {2966} (\bibinfo {year} {2020})}\BibitemShut {NoStop}%
\bibitem [{\citenamefont {Kitsak}\ \emph {et~al.}(2020)\citenamefont {Kitsak}, \citenamefont {Voitalov},\ and\ \citenamefont {Krioukov}}]{Hyperbolic-graph-embeddings-2}%
  \BibitemOpen
  \bibfield  {author} {\bibinfo {author} {\bibfnamefont {M.}~\bibnamefont {Kitsak}}, \bibinfo {author} {\bibfnamefont {I.}~\bibnamefont {Voitalov}},\ and\ \bibinfo {author} {\bibfnamefont {D.}~\bibnamefont {Krioukov}},\ }\bibfield  {title} {\bibinfo {title} {Link prediction with hyperbolic geometry},\ }\href {https://doi.org/10.1103/PhysRevResearch.2.043113} {\bibfield  {journal} {\bibinfo  {journal} {Phys. Rev. Res.}\ }\textbf {\bibinfo {volume} {2}},\ \bibinfo {pages} {043113} (\bibinfo {year} {2020})}\BibitemShut {NoStop}%
\bibitem [{\citenamefont {Faqeeh}\ \emph {et~al.}(2018)\citenamefont {Faqeeh}, \citenamefont {Osat},\ and\ \citenamefont {Radicchi}}]{hyperbolic-graph-embeddings-3}%
  \BibitemOpen
  \bibfield  {author} {\bibinfo {author} {\bibfnamefont {A.}~\bibnamefont {Faqeeh}}, \bibinfo {author} {\bibfnamefont {S.}~\bibnamefont {Osat}},\ and\ \bibinfo {author} {\bibfnamefont {F.}~\bibnamefont {Radicchi}},\ }\bibfield  {title} {\bibinfo {title} {Characterizing the analogy between hyperbolic embedding and community structure of complex networks},\ }\href {https://doi.org/10.1103/PhysRevLett.121.098301} {\bibfield  {journal} {\bibinfo  {journal} {Phys. Rev. Lett.}\ }\textbf {\bibinfo {volume} {121}},\ \bibinfo {pages} {098301} (\bibinfo {year} {2018})}\BibitemShut {NoStop}%
\bibitem [{\citenamefont {Chami}\ \emph {et~al.}(2020)\citenamefont {Chami}, \citenamefont {Wolf}, \citenamefont {Juan}, \citenamefont {Sala}, \citenamefont {Ravi},\ and\ \citenamefont {R{\'e}}}]{hyperbolic-graph-embeddings-4}%
  \BibitemOpen
  \bibfield  {author} {\bibinfo {author} {\bibfnamefont {I.}~\bibnamefont {Chami}}, \bibinfo {author} {\bibfnamefont {A.}~\bibnamefont {Wolf}}, \bibinfo {author} {\bibfnamefont {D.-C.}\ \bibnamefont {Juan}}, \bibinfo {author} {\bibfnamefont {F.}~\bibnamefont {Sala}}, \bibinfo {author} {\bibfnamefont {S.}~\bibnamefont {Ravi}},\ and\ \bibinfo {author} {\bibfnamefont {C.}~\bibnamefont {R{\'e}}},\ }\bibfield  {title} {\bibinfo {title} {Low-dimensional hyperbolic knowledge graph embeddings},\ }in\ \href {https://doi.org/10.18653/v1/2020.acl-main.617} {\emph {\bibinfo {booktitle} {Proceedings of the 58th Annual Meeting of the Association for Computational Linguistics}}}\ (\bibinfo  {publisher} {Association for Computational Linguistics},\ \bibinfo {address} {Online},\ \bibinfo {year} {2020})\ pp.\ \bibinfo {pages} {6901--6914}\BibitemShut {NoStop}%
\bibitem [{\citenamefont {Cohen}()}]{enron-mail}%
  \BibitemOpen
  \bibfield  {author} {\bibinfo {author} {\bibfnamefont {W.}~\bibnamefont {Cohen}},\ }\href@noop {} {\bibinfo {title} {Enron email dataset}},\ \bibinfo {note} {http://www.cs.cmu.edu/~enron/. Accessed in 2009.}\BibitemShut {Stop}%
\bibitem [{\citenamefont {Kumar}\ \emph {et~al.}(2018)\citenamefont {Kumar}, \citenamefont {Hamilton}, \citenamefont {Leskovec},\ and\ \citenamefont {Jurafsky}}]{reddit}%
  \BibitemOpen
  \bibfield  {author} {\bibinfo {author} {\bibfnamefont {S.}~\bibnamefont {Kumar}}, \bibinfo {author} {\bibfnamefont {W.~L.}\ \bibnamefont {Hamilton}}, \bibinfo {author} {\bibfnamefont {J.}~\bibnamefont {Leskovec}},\ and\ \bibinfo {author} {\bibfnamefont {D.}~\bibnamefont {Jurafsky}},\ }\bibfield  {title} {\bibinfo {title} {Community interaction and conflict on the web},\ }in\ \href@noop {} {\emph {\bibinfo {booktitle} {Proceedings of the 2018 World Wide Web Conference on World Wide Web}}}\ (\bibinfo {organization} {International World Wide Web Conferences Steering Committee},\ \bibinfo {year} {2018})\ pp.\ \bibinfo {pages} {933--943}\BibitemShut {NoStop}%
\bibitem [{\citenamefont {Fu}\ and\ \citenamefont {He}(2021)}]{DPPIN}%
  \BibitemOpen
  \bibfield  {author} {\bibinfo {author} {\bibfnamefont {D.}~\bibnamefont {Fu}}\ and\ \bibinfo {author} {\bibfnamefont {J.}~\bibnamefont {He}},\ }\bibfield  {title} {\bibinfo {title} {{DPPIN:} {A} biological repository of dynamic protein-protein interaction network data},\ }\href {https://arxiv.org/abs/2107.02168} {\bibfield  {journal} {\bibinfo  {journal} {CoRR}\ }\textbf {\bibinfo {volume} {abs/2107.02168}} (\bibinfo {year} {2021})},\ \Eprint {https://arxiv.org/abs/2107.02168} {arXiv:2107.02168} \BibitemShut {NoStop}%
\bibitem [{\citenamefont {Rossi}\ and\ \citenamefont {Ahmed}(2015)}]{nr}%
  \BibitemOpen
  \bibfield  {author} {\bibinfo {author} {\bibfnamefont {R.~A.}\ \bibnamefont {Rossi}}\ and\ \bibinfo {author} {\bibfnamefont {N.~K.}\ \bibnamefont {Ahmed}},\ }\bibfield  {title} {\bibinfo {title} {The network data repository with interactive graph analytics and visualization},\ }in\ \href {https://networkrepository.com} {\emph {\bibinfo {booktitle} {AAAI}}}\ (\bibinfo {year} {2015})\BibitemShut {NoStop}%
\bibitem [{\citenamefont {Paranjape}\ \emph {et~al.}(2017)\citenamefont {Paranjape}, \citenamefont {Benson},\ and\ \citenamefont {Leskovec}}]{superuser}%
  \BibitemOpen
  \bibfield  {author} {\bibinfo {author} {\bibfnamefont {A.}~\bibnamefont {Paranjape}}, \bibinfo {author} {\bibfnamefont {A.~R.}\ \bibnamefont {Benson}},\ and\ \bibinfo {author} {\bibfnamefont {J.}~\bibnamefont {Leskovec}},\ }\bibfield  {title} {\bibinfo {title} {Motifs in temporal networks},\ }in\ \href@noop {} {\emph {\bibinfo {booktitle} {Proceedings of the tenth ACM international conference on web search and data mining}}}\ (\bibinfo {year} {2017})\ pp.\ \bibinfo {pages} {601--610}\BibitemShut {NoStop}%
\bibitem [{\citenamefont {Boguna}\ \emph {et~al.}(2021)\citenamefont {Boguna}, \citenamefont {Bonamassa}, \citenamefont {De~Domenico}, \citenamefont {Havlin}, \citenamefont {Krioukov},\ and\ \citenamefont {Serrano}}]{boguna2021network}%
  \BibitemOpen
  \bibfield  {author} {\bibinfo {author} {\bibfnamefont {M.}~\bibnamefont {Boguna}}, \bibinfo {author} {\bibfnamefont {I.}~\bibnamefont {Bonamassa}}, \bibinfo {author} {\bibfnamefont {M.}~\bibnamefont {De~Domenico}}, \bibinfo {author} {\bibfnamefont {S.}~\bibnamefont {Havlin}}, \bibinfo {author} {\bibfnamefont {D.}~\bibnamefont {Krioukov}},\ and\ \bibinfo {author} {\bibfnamefont {M.~{\'A}.}\ \bibnamefont {Serrano}},\ }\bibfield  {title} {\bibinfo {title} {Network geometry},\ }\href@noop {} {\bibfield  {journal} {\bibinfo  {journal} {Nature Reviews Physics}\ }\textbf {\bibinfo {volume} {3}},\ \bibinfo {pages} {114} (\bibinfo {year} {2021})}\BibitemShut {NoStop}%
\bibitem [{\citenamefont {Krioukov}\ \emph {et~al.}(2010)\citenamefont {Krioukov}, \citenamefont {Papadopoulos}, \citenamefont {Kitsak}, \citenamefont {Vahdat},\ and\ \citenamefont {Bogun{\'a}}}]{hyperbolic-complex-network-1}%
  \BibitemOpen
  \bibfield  {author} {\bibinfo {author} {\bibfnamefont {D.}~\bibnamefont {Krioukov}}, \bibinfo {author} {\bibfnamefont {F.}~\bibnamefont {Papadopoulos}}, \bibinfo {author} {\bibfnamefont {M.}~\bibnamefont {Kitsak}}, \bibinfo {author} {\bibfnamefont {A.}~\bibnamefont {Vahdat}},\ and\ \bibinfo {author} {\bibfnamefont {M.}~\bibnamefont {Bogun{\'a}}},\ }\bibfield  {title} {\bibinfo {title} {Hyperbolic geometry of complex networks},\ }\href@noop {} {\bibfield  {journal} {\bibinfo  {journal} {Physical Review E}\ }\textbf {\bibinfo {volume} {82}},\ \bibinfo {pages} {036106} (\bibinfo {year} {2010})}\BibitemShut {NoStop}%
\bibitem [{\citenamefont {Muscoloni}\ \emph {et~al.}(2017)\citenamefont {Muscoloni}, \citenamefont {Thomas}, \citenamefont {Ciucci}, \citenamefont {Bianconi},\ and\ \citenamefont {Cannistraci}}]{hyperbolic-complex-network-2}%
  \BibitemOpen
  \bibfield  {author} {\bibinfo {author} {\bibfnamefont {A.}~\bibnamefont {Muscoloni}}, \bibinfo {author} {\bibfnamefont {J.~M.}\ \bibnamefont {Thomas}}, \bibinfo {author} {\bibfnamefont {S.}~\bibnamefont {Ciucci}}, \bibinfo {author} {\bibfnamefont {G.}~\bibnamefont {Bianconi}},\ and\ \bibinfo {author} {\bibfnamefont {C.~V.}\ \bibnamefont {Cannistraci}},\ }\bibfield  {title} {\bibinfo {title} {Machine learning meets complex networks via coalescent embedding in the hyperbolic space},\ }\href@noop {} {\bibfield  {journal} {\bibinfo  {journal} {Nature communications}\ }\textbf {\bibinfo {volume} {8}},\ \bibinfo {pages} {1615} (\bibinfo {year} {2017})}\BibitemShut {NoStop}%
\bibitem [{\citenamefont {Serafino}\ \emph {et~al.}(2021)\citenamefont {Serafino}, \citenamefont {Cimini}, \citenamefont {Maritan}, \citenamefont {Rinaldo}, \citenamefont {Suweis}, \citenamefont {Banavar},\ and\ \citenamefont {Caldarelli}}]{finite-size-issue-1}%
  \BibitemOpen
  \bibfield  {author} {\bibinfo {author} {\bibfnamefont {M.}~\bibnamefont {Serafino}}, \bibinfo {author} {\bibfnamefont {G.}~\bibnamefont {Cimini}}, \bibinfo {author} {\bibfnamefont {A.}~\bibnamefont {Maritan}}, \bibinfo {author} {\bibfnamefont {A.}~\bibnamefont {Rinaldo}}, \bibinfo {author} {\bibfnamefont {S.}~\bibnamefont {Suweis}}, \bibinfo {author} {\bibfnamefont {J.~R.}\ \bibnamefont {Banavar}},\ and\ \bibinfo {author} {\bibfnamefont {G.}~\bibnamefont {Caldarelli}},\ }\bibfield  {title} {\bibinfo {title} {True scale-free networks hidden by finite size effects},\ }\href@noop {} {\bibfield  {journal} {\bibinfo  {journal} {Proceedings of the National Academy of Sciences}\ }\textbf {\bibinfo {volume} {118}},\ \bibinfo {pages} {e2013825118} (\bibinfo {year} {2021})}\BibitemShut {NoStop}%
\bibitem [{\citenamefont {Voitalov}\ \emph {et~al.}(2019)\citenamefont {Voitalov}, \citenamefont {van~der Hoorn}, \citenamefont {van~der Hofstad},\ and\ \citenamefont {Krioukov}}]{finite-size-issues-2}%
  \BibitemOpen
  \bibfield  {author} {\bibinfo {author} {\bibfnamefont {I.}~\bibnamefont {Voitalov}}, \bibinfo {author} {\bibfnamefont {P.}~\bibnamefont {van~der Hoorn}}, \bibinfo {author} {\bibfnamefont {R.}~\bibnamefont {van~der Hofstad}},\ and\ \bibinfo {author} {\bibfnamefont {D.}~\bibnamefont {Krioukov}},\ }\bibfield  {title} {\bibinfo {title} {Scale-free networks well done},\ }\href {https://doi.org/10.1103/PhysRevResearch.1.033034} {\bibfield  {journal} {\bibinfo  {journal} {Phys. Rev. Res.}\ }\textbf {\bibinfo {volume} {1}},\ \bibinfo {pages} {033034} (\bibinfo {year} {2019})}\BibitemShut {NoStop}%
\bibitem [{\citenamefont {Othmer}\ and\ \citenamefont {Pate}(1980)}]{other-scale-1}%
  \BibitemOpen
  \bibfield  {author} {\bibinfo {author} {\bibfnamefont {H.}~\bibnamefont {Othmer}}\ and\ \bibinfo {author} {\bibfnamefont {E.}~\bibnamefont {Pate}},\ }\bibfield  {title} {\bibinfo {title} {Scale-invariance in reaction-diffusion models of spatial pattern formation.},\ }\href@noop {} {\bibfield  {journal} {\bibinfo  {journal} {Proceedings of the National Academy of Sciences}\ }\textbf {\bibinfo {volume} {77}},\ \bibinfo {pages} {4180} (\bibinfo {year} {1980})}\BibitemShut {NoStop}%
\bibitem [{\citenamefont {Shao}\ \emph {et~al.}(2022)\citenamefont {Shao}, \citenamefont {Li}, \citenamefont {Liu}, \citenamefont {Li}, \citenamefont {Wang}, \citenamefont {Bian}, \citenamefont {Yan}, \citenamefont {Mandrus}, \citenamefont {Liu}, \citenamefont {Zhang} \emph {et~al.}}]{other-scale-2}%
  \BibitemOpen
  \bibfield  {author} {\bibinfo {author} {\bibfnamefont {Z.}~\bibnamefont {Shao}}, \bibinfo {author} {\bibfnamefont {S.}~\bibnamefont {Li}}, \bibinfo {author} {\bibfnamefont {Y.}~\bibnamefont {Liu}}, \bibinfo {author} {\bibfnamefont {Z.}~\bibnamefont {Li}}, \bibinfo {author} {\bibfnamefont {H.}~\bibnamefont {Wang}}, \bibinfo {author} {\bibfnamefont {Q.}~\bibnamefont {Bian}}, \bibinfo {author} {\bibfnamefont {J.}~\bibnamefont {Yan}}, \bibinfo {author} {\bibfnamefont {D.}~\bibnamefont {Mandrus}}, \bibinfo {author} {\bibfnamefont {H.}~\bibnamefont {Liu}}, \bibinfo {author} {\bibfnamefont {P.}~\bibnamefont {Zhang}}, \emph {et~al.},\ }\bibfield  {title} {\bibinfo {title} {Discrete scale invariance of the quasi-bound states at atomic vacancies in a topological material},\ }\href@noop {} {\bibfield  {journal} {\bibinfo  {journal} {Proceedings of the National Academy of Sciences}\ }\textbf {\bibinfo {volume} {119}},\ \bibinfo {pages} {e2204804119} (\bibinfo {year} {2022})}\BibitemShut {NoStop}%
\bibitem [{\citenamefont {Hung}\ \emph {et~al.}(2011)\citenamefont {Hung}, \citenamefont {Zhang}, \citenamefont {Gemelke},\ and\ \citenamefont {Chin}}]{other-scale-3}%
  \BibitemOpen
  \bibfield  {author} {\bibinfo {author} {\bibfnamefont {C.-L.}\ \bibnamefont {Hung}}, \bibinfo {author} {\bibfnamefont {X.}~\bibnamefont {Zhang}}, \bibinfo {author} {\bibfnamefont {N.}~\bibnamefont {Gemelke}},\ and\ \bibinfo {author} {\bibfnamefont {C.}~\bibnamefont {Chin}},\ }\bibfield  {title} {\bibinfo {title} {Observation of scale invariance and universality in two-dimensional bose gases},\ }\href {https://doi.org/10.1038/nature09722} {\bibfield  {journal} {\bibinfo  {journal} {Nature}\ }\textbf {\bibinfo {volume} {470}},\ \bibinfo {pages} {236} (\bibinfo {year} {2011})}\BibitemShut {NoStop}%
\bibitem [{\citenamefont {Vailati}\ \emph {et~al.}(2011)\citenamefont {Vailati}, \citenamefont {Cerbino}, \citenamefont {Mazzoni}, \citenamefont {Takacs}, \citenamefont {Cannell},\ and\ \citenamefont {Giglio}}]{other-scales-4}%
  \BibitemOpen
  \bibfield  {author} {\bibinfo {author} {\bibfnamefont {A.}~\bibnamefont {Vailati}}, \bibinfo {author} {\bibfnamefont {R.}~\bibnamefont {Cerbino}}, \bibinfo {author} {\bibfnamefont {S.}~\bibnamefont {Mazzoni}}, \bibinfo {author} {\bibfnamefont {C.~J.}\ \bibnamefont {Takacs}}, \bibinfo {author} {\bibfnamefont {D.~S.}\ \bibnamefont {Cannell}},\ and\ \bibinfo {author} {\bibfnamefont {M.}~\bibnamefont {Giglio}},\ }\bibfield  {title} {\bibinfo {title} {Fractal fronts of diffusion in microgravity},\ }\href {https://doi.org/10.1038/ncomms1290} {\bibfield  {journal} {\bibinfo  {journal} {Nature Communications}\ }\textbf {\bibinfo {volume} {2}},\ \bibinfo {pages} {290} (\bibinfo {year} {2011})}\BibitemShut {NoStop}%
\bibitem [{\citenamefont {Song}\ \emph {et~al.}(2007)\citenamefont {Song}, \citenamefont {Gallos}, \citenamefont {Havlin},\ and\ \citenamefont {Makse}}]{MEMB}%
  \BibitemOpen
  \bibfield  {author} {\bibinfo {author} {\bibfnamefont {C.}~\bibnamefont {Song}}, \bibinfo {author} {\bibfnamefont {L.~K.}\ \bibnamefont {Gallos}}, \bibinfo {author} {\bibfnamefont {S.}~\bibnamefont {Havlin}},\ and\ \bibinfo {author} {\bibfnamefont {H.~A.}\ \bibnamefont {Makse}},\ }\bibfield  {title} {\bibinfo {title} {How to calculate the fractal dimension of a complex network: the box covering algorithm},\ }\href@noop {} {\bibfield  {journal} {\bibinfo  {journal} {Journal of Statistical Mechanics: Theory and Experiment}\ }\textbf {\bibinfo {volume} {2007}},\ \bibinfo {pages} {P03006} (\bibinfo {year} {2007})}\BibitemShut {NoStop}%
\end{thebibliography}

\clearpage
\newpage

\appendix

\renewcommand{\thefigure}{S\arabic{figure}}
\renewcommand{\thetable}{S\arabic{table}}
\setcounter{figure}{0}
\setcounter{table}{0}

\section{Appendixes}

\begin{figure*}
    \centering
    \includegraphics[width=0.7\textwidth]{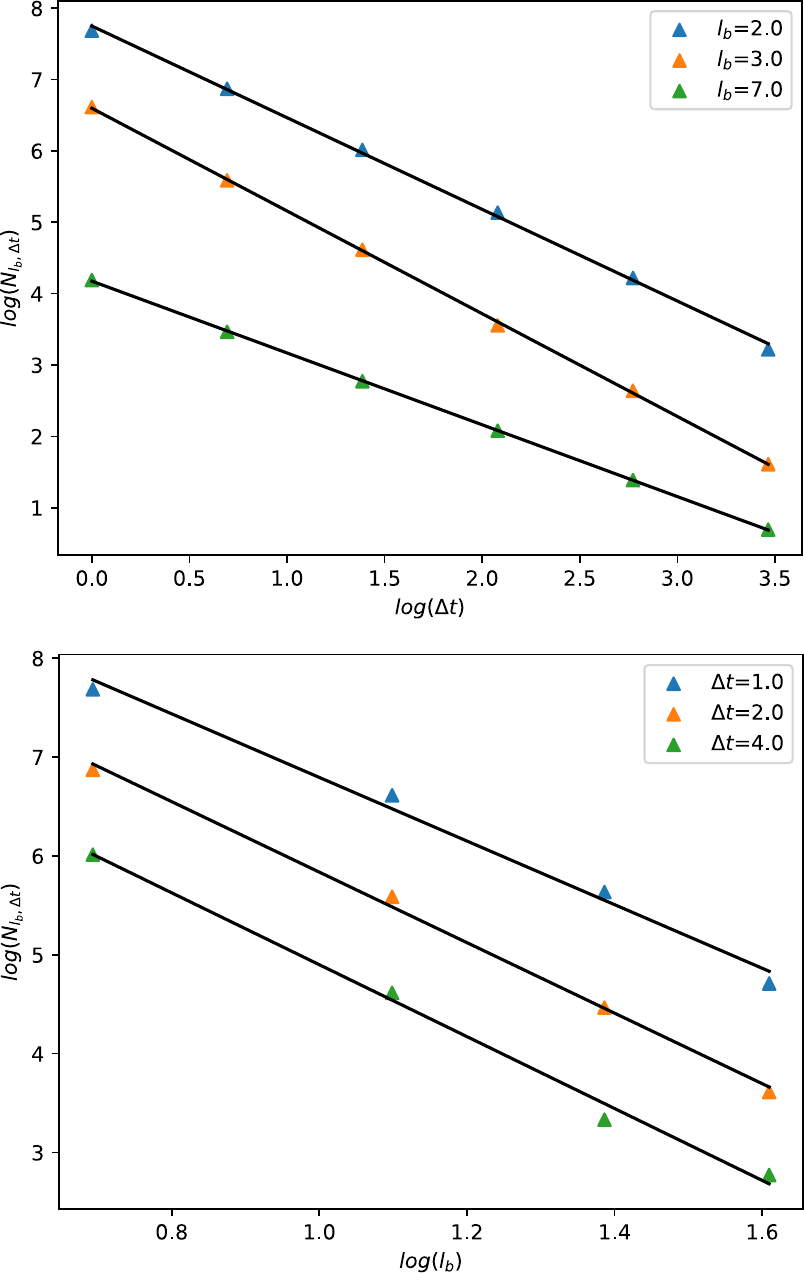}
    \caption{{\bf Scale transformation characteristics of the {\tt ia-email} network}. The top panel shows the variation in the number of boxes with fixed-sized boxes as we change the time-scale. The bottom panel shows the variation of the number of boxes vs the box size for different time-scales.}
    \label{fig:app-1}
\end{figure*}

\begin{figure*}
    \centering
    \includegraphics[width=0.7\textwidth]{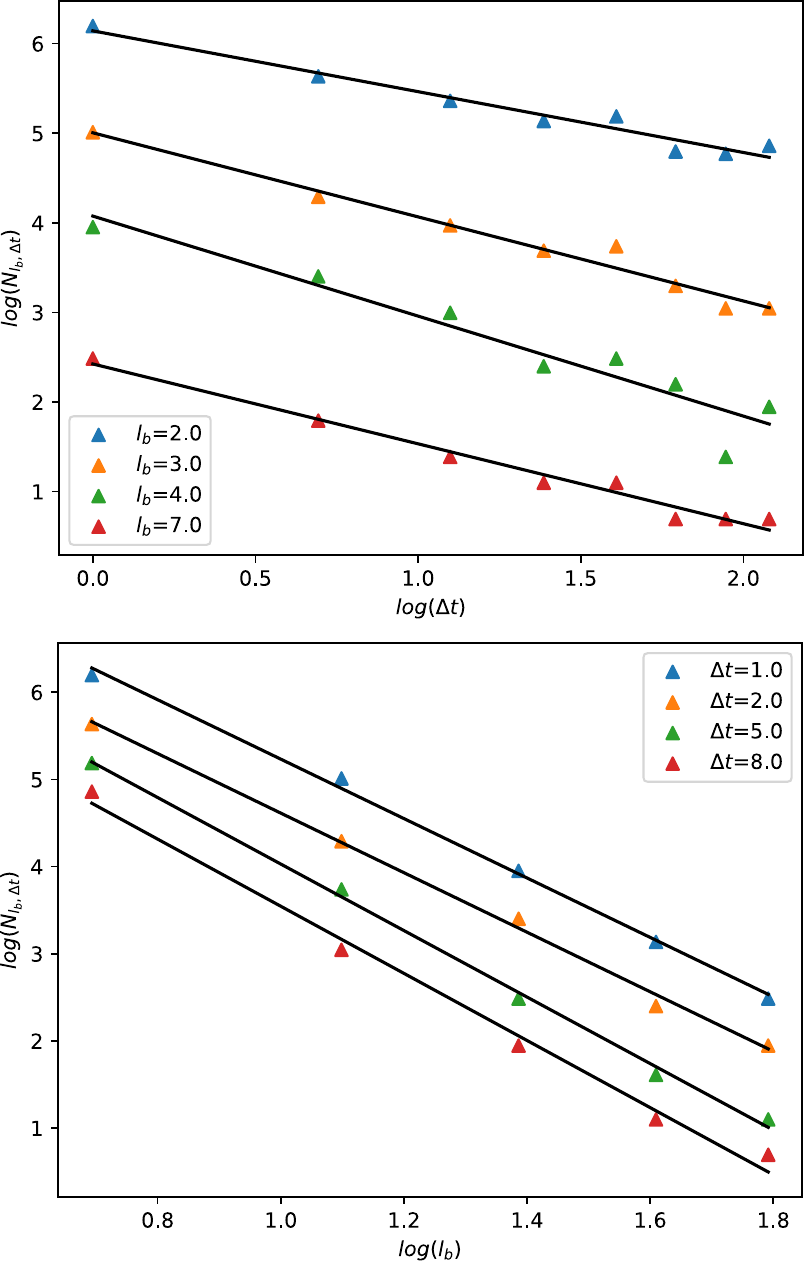}
    \caption{{\bf Scale transformation characteristics of the {\tt ca-cit} network}. The top panel shows the variation in the number of boxes with fixed-sized boxes as we change the time scale. The bottom panel shows the variation of the number of boxes vs the box size for different time scales.}
    \label{fig:app-2}
\end{figure*}

\begin{figure*}
    \centering
    \includegraphics[width=0.7\textwidth]{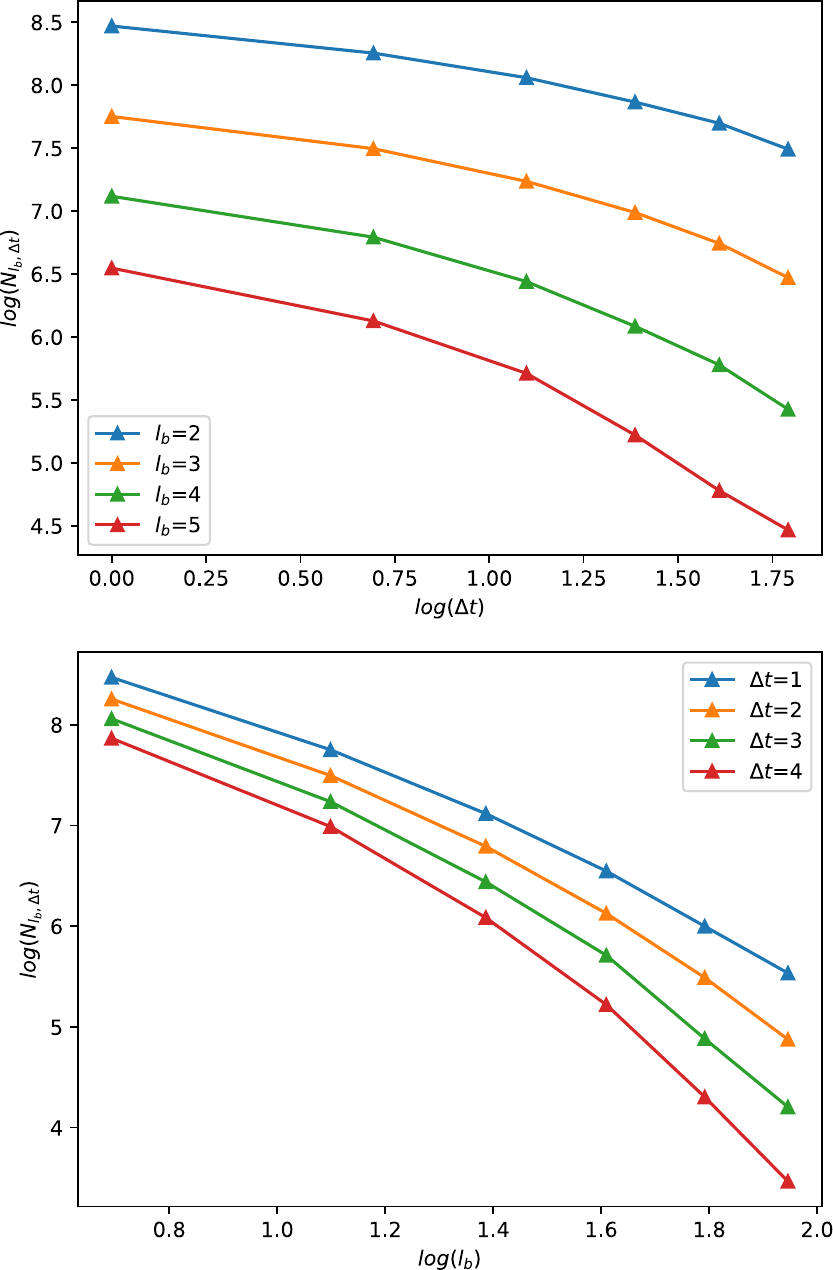}
    \caption{{\bf Scale transformation characteristics of the {\tt DPPI-Babu} network}. The top panel shows the variation in the number of boxes with fixed-sized boxes as we change the time scale. The bottom panel shows the variation of the number of boxes vs the box size for different time scales.}
    \label{fig:app-3}
\end{figure*}

\begin{figure*}
    \centering
    \includegraphics[width=0.7\textwidth]{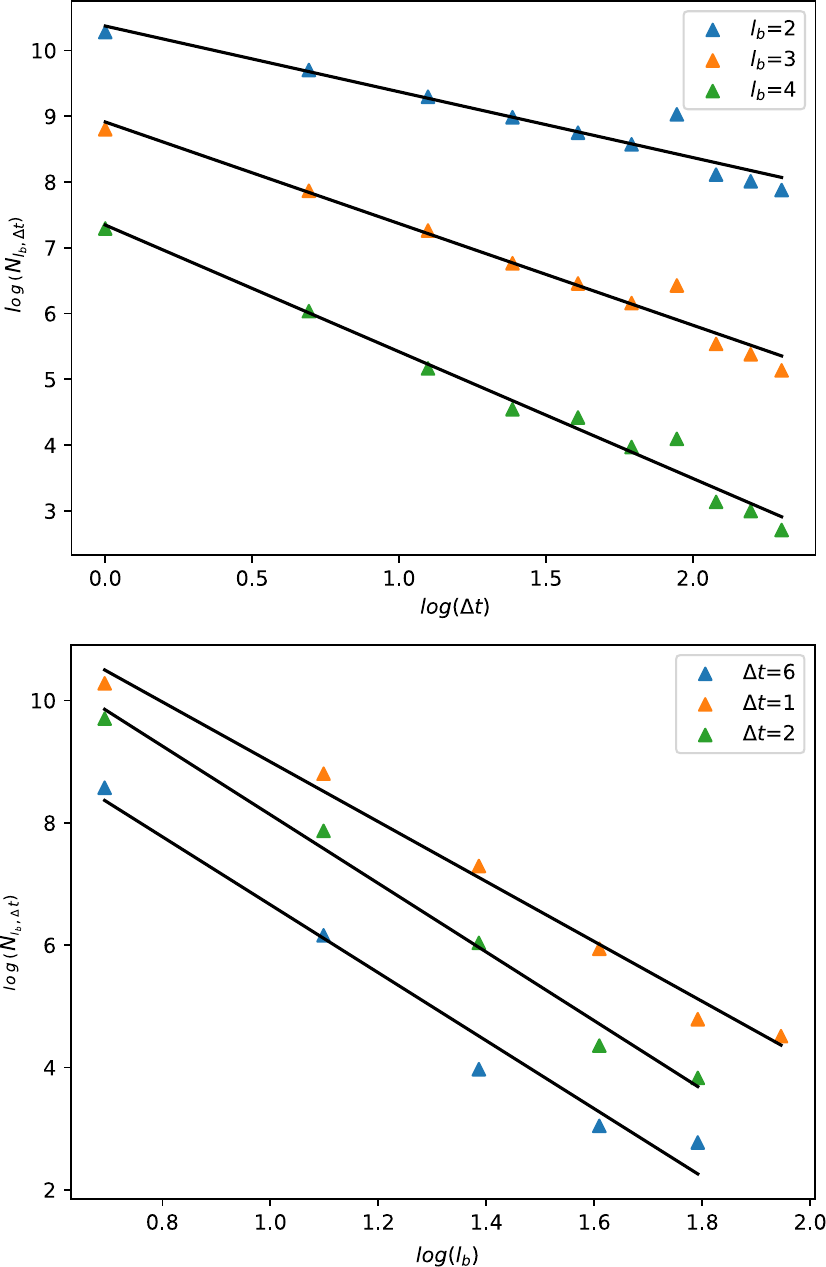}
    \caption{{\bf Scale transformation characteristics of the {\tt reddit-hyperlink} network}. The top panel shows the variation in the number of boxes with fixed-sized boxes as we change the time scale. The bottom panel shows the variation of the number of boxes vs the box size for different time scales.}
    \label{fig:app-4}
\end{figure*}

\begin{figure*}
    \centering
    \includegraphics[width=0.7\textwidth]{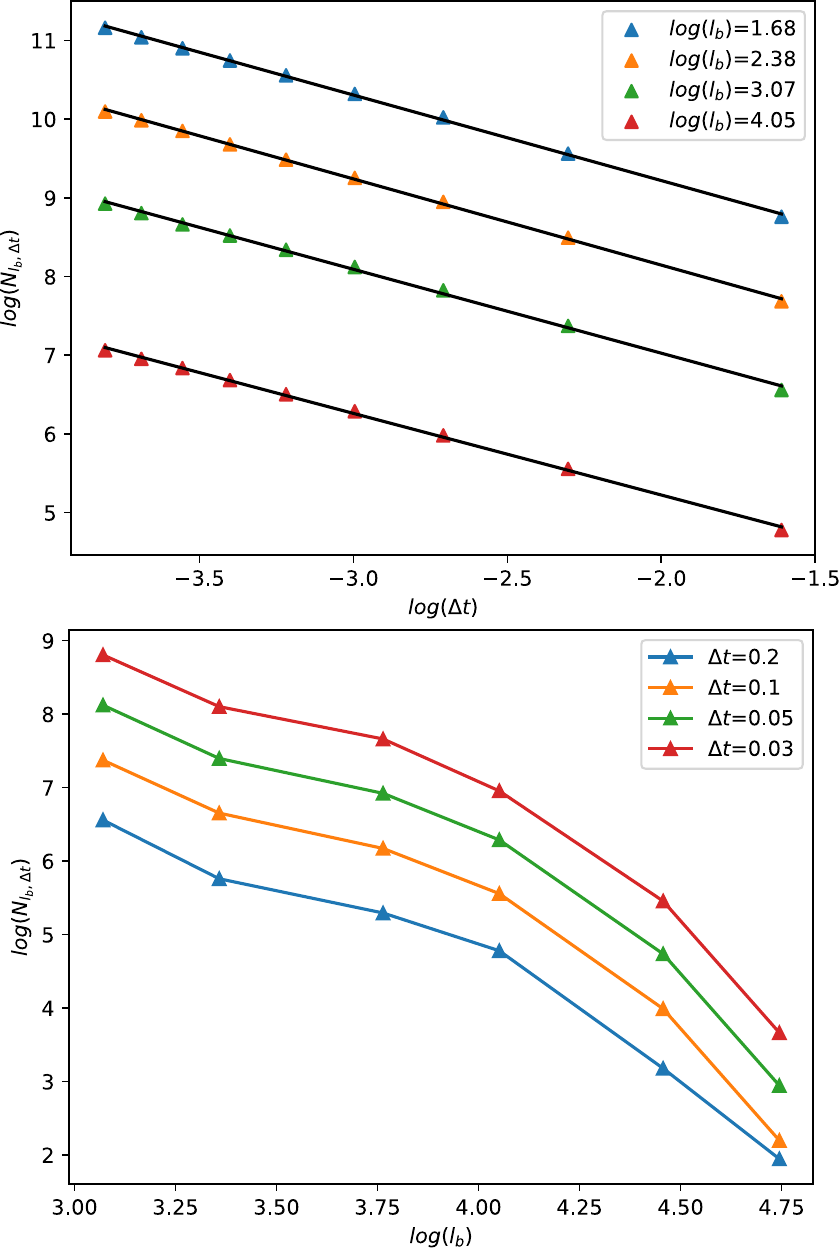}
    \caption{{\bf Scale transformation characteristics of the point-particle system on a hyperbolic space with constant negative curvature}. The top panel shows the variation in the number of boxes with fixed-sized boxes as we change the time scale. The bottom panel shows the variation of the number of boxes vs the box size for different time scales.}
    \label{fig:app-5}
\end{figure*}

\begin{figure*}
    \centering
    \includegraphics[width=0.7\textwidth]{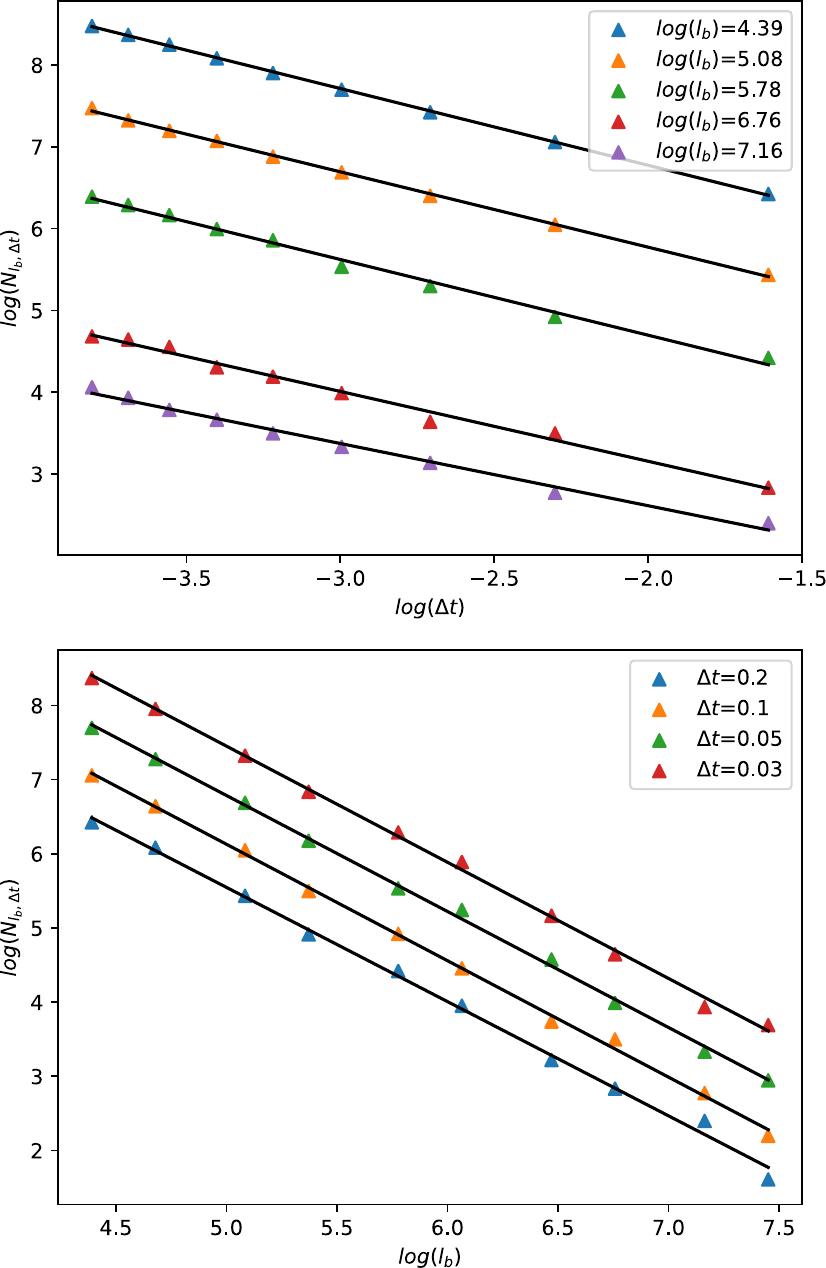}
    \caption{{\bf Scale transformation characteristics of the point-particle system on a hyperbolic space with exponentially increasing negative curvature}. The top panel shows the variation in the number of boxes with fixed-sized boxes as we change the time scale. The bottom panel shows the variation of the number of boxes vs the box size for different time scales.}
    \label{fig:app-6}
\end{figure*}

\noindent{}This document contains additional information related to network datasets and the experimental setup used. In Table~\ref{tab:appendix-table}, we show the timelines and the flow-scale sizes used for experimenting on the different networks. Apart from the flow-scale transformations, we also experiment with the scaling properties of some of the networks when the time scale is changed while keeping the length scales fixed and vice versa. We show these results in Figures~\ref{fig:app-1}, \ref{fig:app-2}, \ref{fig:app-3}, and \ref{fig:app-4}. Finally, we show the same plots in the case of the point-particle systems in Figure~\ref{fig:app-5} (when the curvature of the underlying hyperbolic geometry is constant) and Figure~\ref{fig:app-6} (when the curvature increases exponentially).

\begin{table*}[!t]
\small
    \centering
    \caption{{\bf Choices of scale-sizes for different datasets.} $\Delta t$ denotes the time-scale size that will be multiplied by $k$ for different time-scale renormalizations. $r_b$ denotes the box radius in MEMB. $t_0$ and $t_N$ denote the start and end times of interactions in the respective networks, respectively. For {\tt DPPI-Babu}, the unit of time is as specified in the original corpus.}
    \begin{tabular}{|c|c|c|c|c|}
    \hline
        Network & $\Delta t$ & $t_0$ & $t_N$ & Choices of $(k, r_b)$\\
    \hline
      {\tt ia-email}  & 30 days & September 18, 1999 & September 7, 2001 &  (0.5, 1), (1., 2), (1.5, 3), (2, 4), (2.5, 5), (3, 6), (4, 8)\\
      {\tt reddit-hyperlink}  & 30 days & December 31, 2013 & December 21, 2015 & (0.5, 1), (1., 2), (1.5, 3), (2, 4), (2.5, 5), (3, 6)\\
      {\tt DPPI-Babu}  & 1 & 0 & 36 & (1, 1), (2, 2), (3, 3), (4, 4), (5,5), (6, 6), (7,7)\\
      {\tt ca-cit}  & 90 days & February 27, 1995 & January 26, 2001 & (0.5, 1), (1., 2), (1.5, 3), (2, 4), (2.5, 5)\\
      {\tt superuser}  & 30 days & March 12, 2010 & February 9, 2016 & (0.5, 1), (1., 2), (1.5, 3), (2, 4), (2.5, 5) \\
      {\tt wiki-talk}  & 30 days & October 16, 2002 & September 30, 2005 & (0.5, 1), (1, 2), (1.5, 3), (2, 4), (2.5, 5)\\
    \hline
    \end{tabular}
    \label{tab:appendix-table}
\end{table*}

\end{document}